\documentclass[twocolumn,superscriptaddress,pre,floatfix]{revtex4} 

\usepackage{graphicx}
\usepackage{bm}	
\usepackage{color}
\usepackage{xcolor}
\usepackage{epsfig}
\usepackage{amsmath}
\usepackage{amssymb}
\usepackage{longtable}
\usepackage{float}
\usepackage{mathtools}
\usepackage{xparse}
\usepackage{hyperref}
\usepackage{times}
\usepackage[normalem]{ulem}

\usepackage{sidecap}
\sidecaptionvpos{figure}{t}

\newcommand{\be}{\begin{equation}}
\newcommand{\ee}{\end{equation}}
\newcommand{\bd}{\begin{displaymath}}
\newcommand{\ed}{\end{displaymath}}
\newcommand{\BE}{\begin{eqnarray}}
\newcommand{\EE}{\end{eqnarray}}

\newcommand{\Jferro}{J_{\mathrm{F}}}

\newcommand{\Ham}[1]{H_{\mathrm{#1}}}
\newcommand{\Hnumfault}{\Ham{numfaults}}
\newcommand{\Hgate}{\Ham{gate}}
\newcommand{\Hfaultset}{\Ham{faultset}}

\newcommand{\Hmultfault}{\Ham{multfault}}
\newcommand{\weight}[1]{\lambda_{\mathrm{#1}}}
\newcommand{\faultweight}{\weight{faultset}}
\newcommand{\multfaultweight}{\weight{multfault}}
\newcommand{\gateweight}{\weight{gate}}
\newcommand{\OR}{\textsc{OR}}
\newcommand{\AND}{\textsc{AND}}
\newcommand{\XOR}{\textsc{XOR}}
\newcommand{\EQ}{\textsc{EQ}}
\newcommand{\BUFFER}{\textsc{BUFFER}}
\newcommand{\NOR}{\textsc{NOR}}
\newcommand{\NOT}{\textsc{NOT}}
\newcommand{\NAND}{\textsc{NAND}}

\definecolor{darkgreen}{rgb}{0.0, 0.5, 0.0}

\providecommand{\abs}[1]{\lvert#1\rvert}

\NewDocumentCommand{\ceil}{s O{} m}{
  \IfBooleanTF{#1} 
    {\left\lceil#3\right\rceil} 
    {#2\lceil#3#2\rceil} 
}

\begin{document}

\title{Readiness of Quantum Optimization Machines for Industrial
Applications}

\author{Alejandro Perdomo-Ortiz}
\email{alejandro@zapatacomputing.com}
\affiliation{Quantum Artificial Intelligence Lab., NASA Ames Research Center, Moffett Field, California 94035, USA}
\affiliation{USRA Research Institute for Advanced Computer Science (RIACS), Mountain View California 94043, USA}
\affiliation{Zapata Computing Inc., 439 University Avenue, Office 535, Toronto, ON, M5G 1Y8}
\affiliation{Department of Computer Science, University College London, WC1E 6BT London, UK}

\author{Alexander Feldman}
\affiliation{Palo Alto Research Center, 3333 Coyote Hill Road, Palo Alto, California 94304, USA}

\author{Asier Ozaeta}
\affiliation{QC Ware Corp., 125 University Ave., Suite 260, Palo Alto, California 94301, USA}

\author{Sergei V. Isakov}
\affiliation{Google Inc., 8002 Zurich, Switzerland}

\author{Zheng Zhu}
\affiliation{Department of Physics and Astronomy, Texas A\&M University, College Station, Texas 77843-4242, USA} 

\author{Bryan O'Gorman}
\affiliation{Quantum Artificial Intelligence Lab., NASA Ames Research Center, Moffett Field, California 94035, USA}
\affiliation{Berkeley Center for Quantum Information and Computation, Berkeley, California 94720 USA}
\affiliation{Department of Chemistry, University of California, Berkeley, California 94720 USA}

\author{Helmut G.~Katzgraber}
\affiliation{Department of Physics and Astronomy, Texas A\&M University, College Station, Texas 77843-4242, USA} 
\affiliation{1QB Information Technologies (1QBit), Vancouver, British Columbia, Canada V6B 4W4}
\affiliation{Santa Fe Institute, 1399 Hyde Park Road, Santa Fe, New Mexico 87501, USA}

\author{Alexander Diedrich}
\affiliation{Fraunhofer IOSB-INA, Lemgo, Germany}

\author{Hartmut Neven}
\affiliation{Google Inc., Venice, California 90291, USA}

\author{Johan de Kleer}
\affiliation{Palo Alto Research Center, 3333 Coyote Hill Road, Palo Alto, California 94304, USA}

\author{Brad Lackey}
\affiliation{Joint Center for Quantum Information and Computer Science, University of Maryland, College Park, Maryland 20742, USA}
\affiliation{Departments of Computer Science and Mathematics, University of Maryland, College Park, Maryland 20742, USA}
\affiliation{Mathematics Research Group, National Security Agency, Ft. George G. Meade, Maryland 20755, USA}

\author{Rupak Biswas}
\affiliation{Exploration Technology Directorate, NASA Ames Research Center, Moffett Field, California 94035, USA}

\date{\today}

\begin{abstract}
There have been multiple attempts to demonstrate that quantum annealing and, in particular, quantum annealing on quantum annealing machines, has the potential to outperform current classical optimization algorithms
implemented on CMOS technologies. The benchmarking of these devices has been controversial. Initially, random spin-glass problems were used, however, these were quickly shown to be not well suited to detect any
quantum speedup. Subsequently, benchmarking shifted to carefully crafted
synthetic problems designed to highlight the quantum nature of the
hardware while (often) ensuring that classical optimization techniques
do not perform well on them. Even worse, to date a true sign of improved
scaling with the number of problem variables remains elusive when compared
to classical optimization techniques.  Here, we analyze the readiness of
quantum annealing machines for real-world application problems. These
are typically not random and have an underlying structure that is hard
to capture in synthetic benchmarks, thus posing unexpected challenges
for optimization techniques, both classical and quantum alike. We
present a comprehensive computational scaling analysis of fault diagnosis in digital circuits, considering architectures beyond D-wave quantum annealers. We find that the instances generated from
real data in multiplier circuits are harder than other representative
random spin-glass benchmarks with a comparable number of variables.
Although our results show that transverse-field quantum annealing is
outperformed by state-of-the-art classical optimization algorithms,
these benchmark instances are hard and small in the size of the input,
therefore representing the first industrial application ideally suited
for testing near-term quantum annealers and other quantum algorithmic strategies for optimization problems.
\end{abstract}

\maketitle

\section{Introduction}

Quantum annealing (QA)
\cite{kadowaki:98,finnila:94,farhi:01,santoro:02,das:05,santoro:06,das:08}
has been proposed as the most natural quantum-computing framework to
tackle combinatorial optimization problems, where finding the
configuration that minimizes an application-specific cost function is at
the core of the computational task. Despite multiple studies
\cite{johnson:11,dickson:13,boixo:13a,katzgraber:14,boixo:14,ronnow:14a,katzgraber:15,heim:15,hen:15a,rieffel:15,boixo:16,denchev:16,king:19},
a definite detection of quantum speedup \cite{ronnow:14a,mandra:16b}
remains elusive. Random spin-glass benchmarks \cite{ronnow:14a} have
been shown to be deficient in the detection of quantum speedup
\cite{katzgraber:14,katzgraber:15}, which is why the community has
shifted to carefully crafted synthetic benchmarks
\cite{denchev:16,king:19}. While these have shown that QA
has a constant speedup over state-of-the-art classical optimization
techniques, their value for real-world applications remains
controversial.

Although the first proposal for a QA implementing
combinatorial optimization problems with real constrains as they appear
in real-world application was proposed close to a decade ago
\cite{perdomo08}, the question of whether a quantum annealer can have a
quantum speedup on any real-world applications remains an open one.
From the many applications implemented in quantum annealers (see, for
example,
Refs.~\cite{PerdomoOrtiz2012_LPF,Gaitan2012,OGormanEPJST2015,PerdomoOrtiz_EPJST2015,rieffel:15,
Zick2015, Neukart2017}), fault diagnosis has been one of the leading
candidates to benchmark the performance of D-Wave devices as optimizers
\cite{PerdomoOrtiz_EPJST2015,Bian2016}. From the range of circuit
model-based fault-diagnosis problems \cite{feldman10safari} we restrict
our attention here to \textit{combinational circuit fault diagnosis}
(CCFD), which in contrast to sequential circuits, does not have any
memory components and the output is entirely determined by the present
inputs.

Using CCFDs, we illustrate the challenges and the readiness of quantum
annealers for solving real-world problems by providing a comprehensive
computational scaling analysis of a real-world application. We compare
quantum Monte Carlo (QMC) simulations and QA experiments
on the D-Wave Systems Inc.~D-Wave 2X quantum annealer to several
state-of-the-art classical solvers on conventional computer hardware. 
More specifically, our work is motivated by these open questions in quantum optimization with QA hardware:

\medskip

\noindent 1.~What is the payoff of investing in the construction of
specialized quantum hardware that natively matches the connectivity and
interactions (e.g., many-body terms in higher-order Hamiltonians)
dictated by the cost function of an actual application?

\medskip

\noindent 2.~What could be the impact in the computational scaling of
different annealing schedules or the addition of more complex driver,
such as nonstoquastic Hamiltonians?

\medskip

\noindent 3.~Does quantum Monte Carlo reproduce the computational scaling
of the current generation of D-Wave QA machines?

\medskip

\noindent Keeping in mind these are very general and ambitious goals for a single work like the one presented here, we focus our scope only to the case of optimization instances generated from these real-world scenarios. We discuss the importance of each of these algorithmic and
architectural design aspects related to each of the questions above,
from an application-centric and physics-focused perspective, providing answers or insights only in some cases and under the assumptions and computational resources described throughout this work. It is
demonstrated that CCFD instances based on Boolean multiplier circuits
are harder than other representative random spin-glass benchmarks. This
makes the diagnosis of Boolean multipliers a prime application for
benchmarking QA architectures.  Since our work hints at the need for further developments, with the inclusion of more powerful driver Hamiltonians among one of the interesting research directions in the search for quantum advantage, CCFD instances are ideal industrial application problems for testing
such incremental improvements in near-term quantum annealers and novel
quantum algorithmic strategies for optimization problems.

Although tangential to the key results in this paper, in Appendix~\ref{s:dw2x_vs_qmc} we discuss the last of the three questions above. The main reason for including this section is to highlight that from our perspective of the first scaling analysis of a real-world application, our results indicate that given the hardness of our instances compared to synthetic data sets, the scaling becomes a moot question. This is, even assuming a favorable scenario where the simulated quantum-annealing (SQA) scaling slope matches the D-Wave 2X device (DW2X) scaling, the prefactor is large enough that attempting to use computational resources for simulating SQA becomes prohibitedly expensive. This has not been the case with other studies on synthetic instances~\cite{denchev:16}.


\section{Benchmark problem}
\label{sec:problem}

\begin{figure*}
\includegraphics[width=.80\textwidth]{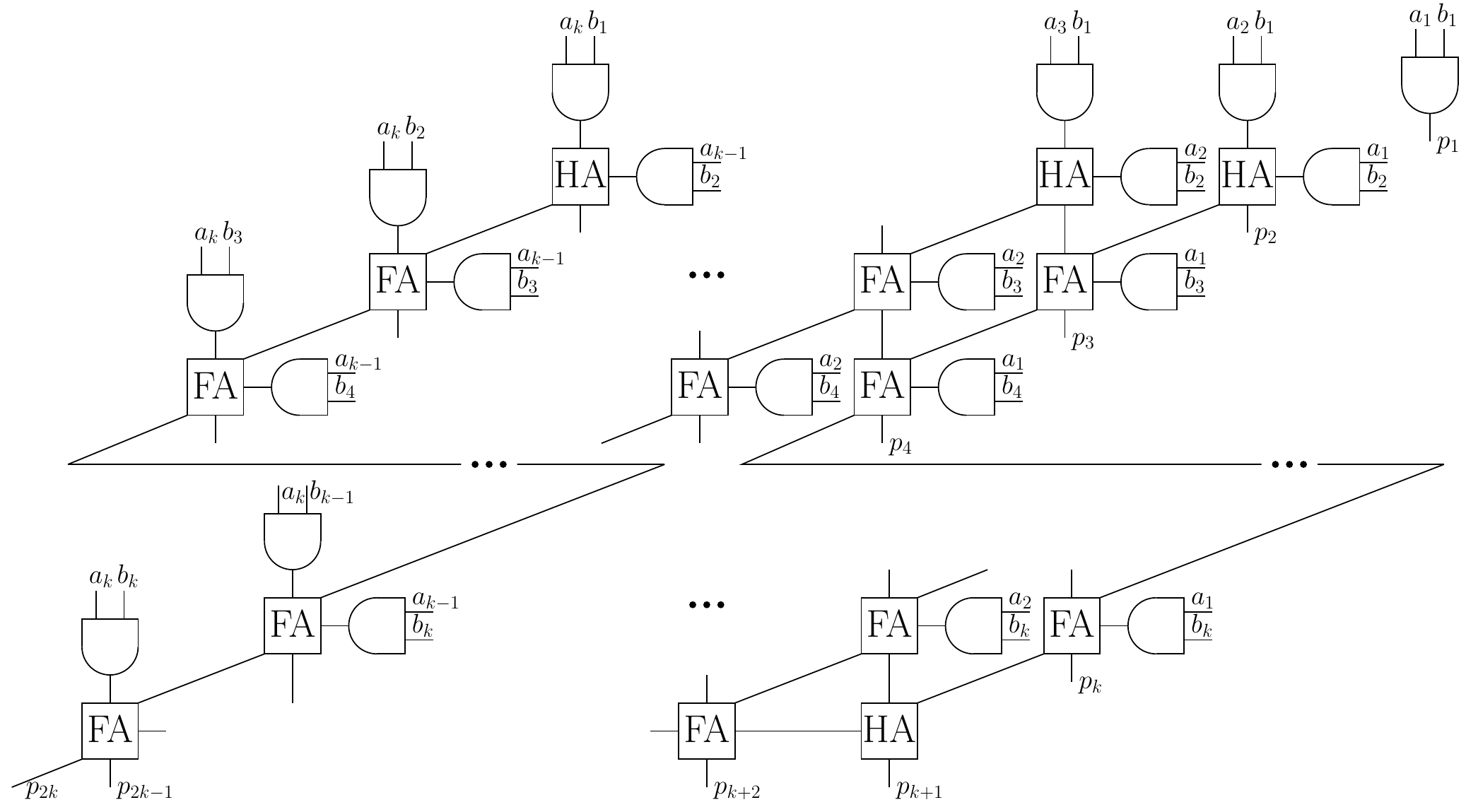}
\caption{
Multiplier circuits used to generate our CCFD benchmark instances. In
this example the multiplication of two numbers represented as $k$-digit
binary numbers, $a_1 a_2 \cdots a_k$ and $b_1 b_2 \cdots b_k$ is shown,
resulting in a product output of length $2k$, corresponding to $p_1 p_2
\cdots p_{2k}$. HA and FA denote half-adder and full-adder circuit
modules, respectively.
}\label{f:ccfd_multk-k}
\end{figure*}

To benchmark quantum annealers with different physical hardware
specifications, we generate a family of multiplier circuits of varying
size. The circuit size is determined by the size of two binary numbers
of bit lengths $n$ and $m$, respectively, to be multiplied. Figure
\ref{f:ccfd_multk-k} illustrates the layout of the multiplication
circuit for two binary numbers, each of length $k$.

\begin{figure}[!h]
\includegraphics[width=.45\textwidth]{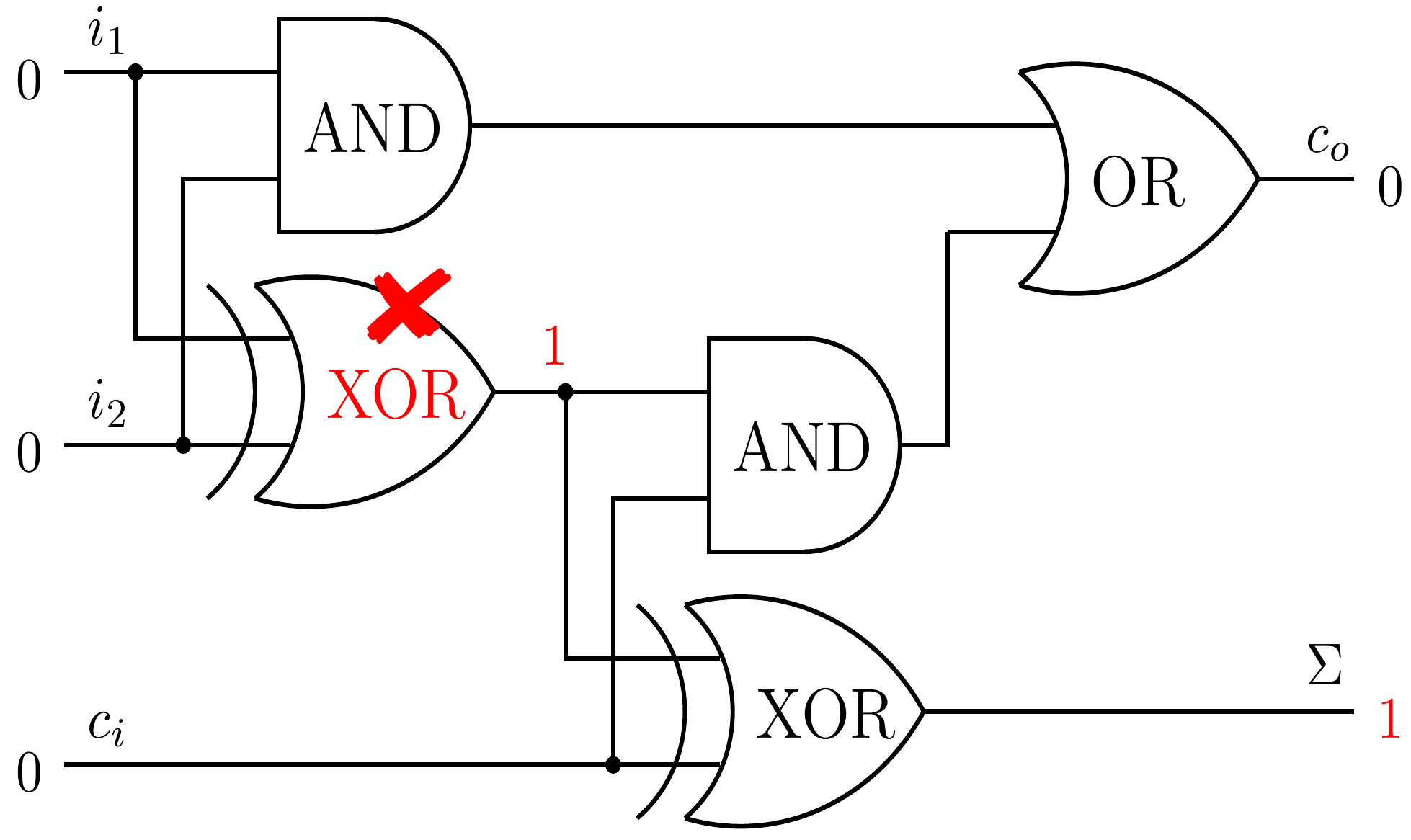}
\caption{
Example of a model-based fault diagnosis in combinational circuits
(CCFD) on a small full adder circuit. In this work, the CCFD
optimization problem consists in finding the smallest set of Boolean gate
outputs that, when stuck-at-one, match an input-output observation
vector. This setting where one restricts the expected fault behavior of
the gates is known as the \textit{strong-fault model} (see, for example,
Ref.~\cite{Feldman2007}). Although in principle each gate can have a
characteristic fault mode, without loss of generality, we adopt
stuck-at-one as the fault mode for all gates. Generalizations to other
common fault modes and multiple fault modes per gate are detailed in
Appendix~\ref{s:mapping}. In this example, the flagged XOR gate is
faulty, because its nominal behavior should yield an output equal to
zero.  The diagnosis explains the input $\{i_1 = 0,i_2= 0, c_i =0\}$ and
the apparently anomalous output $\{c_0 =0, \Sigma =1\}$.
}
\label{f:ccfd_example} 
\end{figure}

The optimization problem consists in diagnosing the health status of
each of the gates in the circuit, given an observation vector consisting
of inputs and outputs, as illustrated in Fig.~\ref{f:ccfd_example}. For
the generation of the problem instances, we focus on problems where the
output is not consistent with the multiplication of the two input
numbers and therefore the system is expected to have at least one fault.
Under the assumption that all the gates have the same failure
probability, the problem of finding the most probable diagnosis is
reduced to finding the valid diagnoses with the minimal number of
faults (see Appendix \ref{s:mapping} for details). It is important to note that all CCFD instances used in this study were randomly generated by injecting a number of faults equivalent to the number of outputs in the circuit [$(n+m)$ for a ($n$-bit) $\times (m$-bit) multiplier circuit]. After the random fault injection of cardinality $(n+m)$, a random input is generated and the corresponding output is obtained by propagation of the input under the corresponding fault injection. Hence, we guarantee that every random input-output pair generated this way has at least one solution. The simpler strategy of generating  random input-output vectors can lead to problems that do not have a solution under the diagnosis model. In the case of instances with many valid minimal solutions, we count all the ones found by the stochastic algorithms in the estimation of the success probability.

From a computational complexity perspective, the CCFD problem is non-deterministic polynomial-time hard (NP-hard)
\cite{eiter95complexity}, and it corresponds to the minimization task we
aim to solve either with QA on the DW2X at NASA, a continuous-time version \cite{rieger:kawashima} of SQA \cite{santoro:06,boixo:14,ronnow:14a}
as a QMC-based solver, or other classical optimization techniques, such
as simulated annealing (SA) \cite{kirkpatrick:83,isakov:15}, parallel
tempering Monte Carlo (modified as a solver)
\cite{hukushima:96,katzgraber:06a,moreno:03} combined with isoenergetic
cluster updates \cite{zhu:15b} (PTICM), or current specialized SAT-based
solvers tailored for this CCFD problem described in Appendix
\ref{s:methods}.

To perform a scaling analysis it is key to be able to generate a data
set with varying input size and with a high intrinsic hardness such that
classical solvers have a harder time, increasing the chances that our
instances fall into the hard asymptotic regime for both classical and
quantum approaches. This has been one of the challenges for benchmarking
early QA devices, where the first proposals
\cite{boixo:14,ronnow:14a} were convenient but turned out to be too easy
for benchmarking purposes \cite{katzgraber:14}. More recently,
benchmarking has focused on carefully designed synthetic problems
\cite{denchev:16,hen:15a,mandra:16,king:19}. However, as we demonstrate in
Sec.~\ref{ss:hardness}, CCFD-based problems are the hardest benchmarking
problems currently available.

\section{Results and Discussion}\label{s:results}

\subsection{Benchmarking real-world applications}~\label{s:bencmarking_ccfd}

\begin{SCfigure*}
\includegraphics[width=0.48\textwidth]{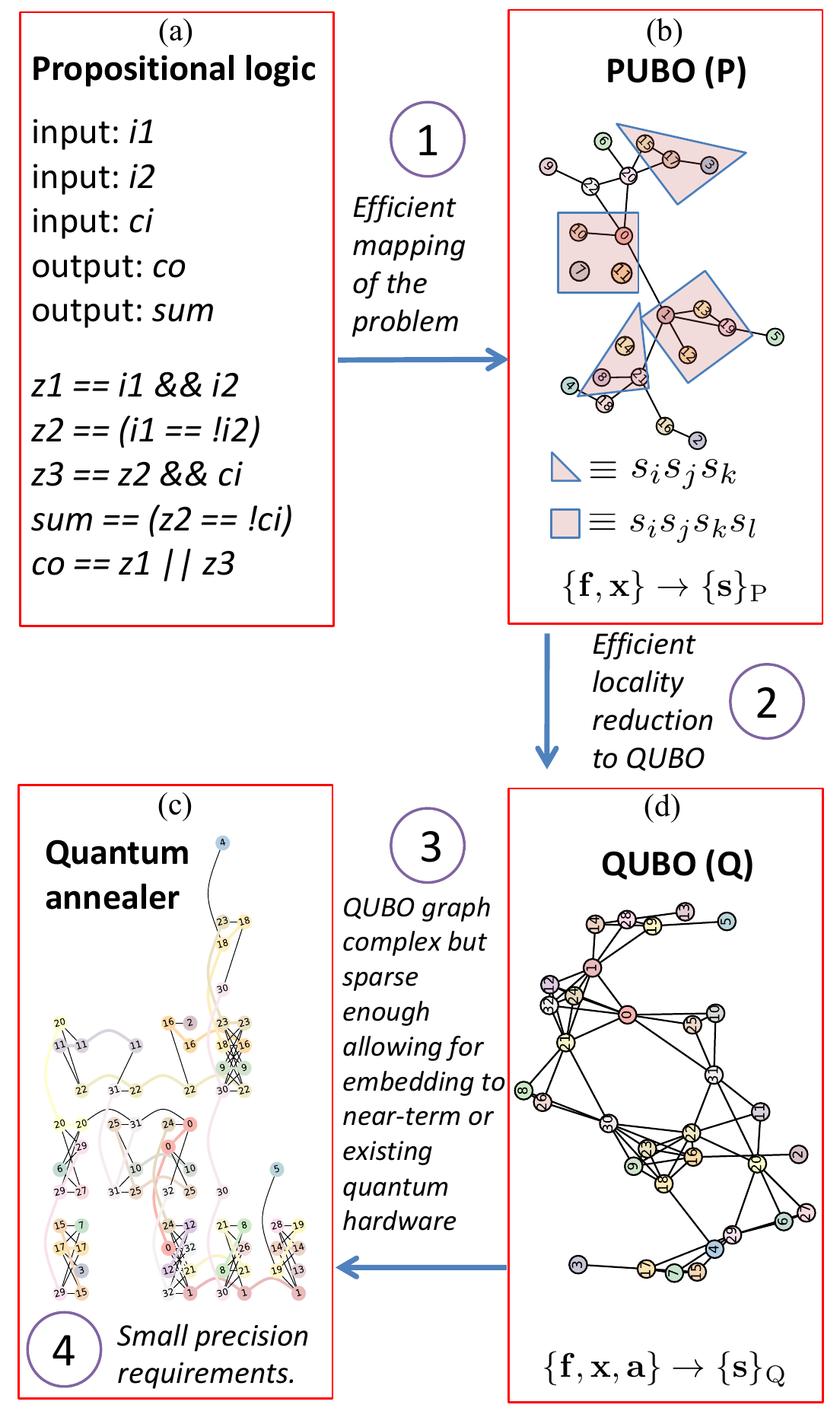}
\label{f:ccfd-benchmark}
\hspace*{1.0em}
\caption{
Generation of benchmarks driven with real-world scenarios flowchart. The
first challenge consists in finding an efficient translation of the
natural description of the optimization problem into a pseudo-Boolean
function [panel (a)], $H_{\rm{P}}(\mathbf{s}_{\rm{P}})$, with domain
$\mathbf{s}_{\rm{P}} \in \{+1,-1\}^{N_{\rm{P}}}$ and co-domain in
$\mathbb{R}$. The resulting PUBO problem consists in finding assignments $\mathbf{s}^{*}_{\rm{P}}$
which minimizes the quartic-degree polynomial
$H_{\rm{P}}(\mathbf{s}_{\rm{P}})$ [panel (b)].  This specific degree,
known as the \textit{locality} of the Hamiltonian (here, 4-local),
arises from the effective local interactions between gates with two
input variables $x_{i,1}$ and $x_{i,2}$, the wire output $z_i$, and the
health variable, $f_i$, associated to each gate. By adding ancilla qubits, ($\mathbf{a}$), the quartic (4-local) $s_i s_j s_k
s_l$ and cubic (3-local) $s_i s_j s_k$ terms can be reduced to an
effective 2-local Hamiltonian defining an effective quadratic
unconstrained binary optimization (QUBO) version of the problem instance
[panel (d)]. Finally, minor embedding can be used to embed the QUBO
into the physical hardware -- in this case the chimera structure (see
Fig.~\ref{fig:chimera}) of the D-Wave quantum annealers [panel (c)].
The cost of the embedding is an additional overhead in the number of
qubits.  While the ``propositional logic'' panel contains the
description of the full-adder circuit in Fig.~\ref{f:ccfd_example}, the
remaining are realistic representations of one of the smallest instances
from our multiplier circuit with $23$, $33$, and $72$ qubits (or spin
variables), for its PUBO, QUBO, and DW2X representation, respectively.
In this work, we assess the impact on the performance of each of these
representations and also perform experiments on the DW2X. Steps 1 -- 4
denote some of the desiderata for an application to be a potential
candidate for benchmarking next generation of quantum annealers.  Note
that while the D-Wave device requires the embedding of a QUBO. However,
future hardware implementations might include $k$-local interactions
with $k > 2$. Therefore, we perform the classical simulations both in
the QUBO and PUBO representations to compare both approaches.}
\end{SCfigure*}

Figure \ref{f:ccfd-benchmark} summarizes the main challenges when
benchmarking applications with QA devices. The first step
consists of translating the standard format describing the rules and
constrains of the minimization problem into a pseudo-Boolean polynomial
function $H_{\rm{P}}(\mathbf{s}_{\rm{P}})$, with domain
$\mathbf{s}_{\rm{P}} \in \{+1,-1\}^{N_{\rm{P}}}$ and co-domain in
$\mathbb{R}$. Appendix~\ref{s:mapping} details the construction of
$H_{\rm{P}}(\mathbf{s}_{\rm{P}})$ for this problem of minimal fault
diagnosis in combinational circuits. The task to be solved consists in
finding, within the search space with $2^{N_{\rm{P}}}$ possible
solutions, the assignment $\mathbf{s^{*}_{\rm{P}}}$ that minimizes
$H_{\rm{P}}(\mathbf{s}_{\rm{P}})$. Because the pseudo-Boolean function
is a polynomial expression in the binary variables
$\mathbf{s}_{\rm{P}}$, this optimization problem is known as a PUBO
problem which stands for \textit{polynomial unconstrained binary
optimization} problem. Note that sometimes these problems are also
referred to as HOBOs, i.e, \textit{higher-order binary optimization}
problem. The specific case of a quadratic function leads to the known quadratic unconstrained binary optimization (QUBO) \cite{lucas:14} which is the type that is natively implemented in
D-Wave quantum annealers. See Appendix.~\ref{s:qa} for more details on
the QA implementation.

In the case of the benchmark of multiplier circuits, the standard
problem description format is a list of propositional logic formulas
similar to the ones given in Fig.~\ref{f:ccfd-benchmark}, corresponding
to the nominal behavior of each gate within the full-adder circuit
illustrated in Fig.~\ref{f:ccfd_example}. For the case of the
strong-fault model \cite{Feldman2007} considered here, one needs to add
specific propositional logic formulas associated with the expected
behavior when each gate is faulty. Without loss of generality, and for
the purpose of the benchmark generated here, we considered that whenever
any gate fails, it would be in a \textit{stuck-at-one} mode or
equivalently, in propositional logic, $f_i \Rightarrow z_i$. Here $f_i$
denotes the health variable associated with the $i$th gate and $z_i$
its corresponding gate output. Note that $f_i =1$ means faulty and
$f_i=0$ nominal. Extensions to other fault modes are described in
Appendix.~\ref{s:mapping}. In the specific mapping considered here,
$\mathbf{s}_{\rm{P}}$ contains the health variables, along with
variables specifying the values for each of the internal wires within
the multiplier circuit.

A generic classical solver such as SA or PTICM can tackle the
optimization problem in th PUBO representation directly because one can
easily evaluate $H_{\rm{P}}(\mathbf{s}_{\rm{P}})$. As shown in
Sec.~\ref{ss:physics-ccfd-scaling}, working in this PUBO representation
is the preferred approach from the application perspective. As mentioned
in the explicit mapping construction in Sec.~\ref{s:mapping},
$H_{\rm{P}}(\mathbf{s}_{\rm{P}})$ is a polynomial with at most quartic
degree, independent of the circuit size. A quantum annealer capable of
implementing such quartic polynomials can certainly aim at solving the
problem in this representation. Given the possibility of such
experimental designs (see, for example, Ref.~\cite{Chancellor2017}), we
also consider hypothetical quantum annealers that we study using SQA to
assess the impact in the performance of working with a quantum annealer
that can natively solve the PUBO problem. Unfortunately, no such devices
exist to date and there is an overhead in representing the quartic
(4-local) and cubic (3-local) monomials in
$H_{\rm{P}}(\mathbf{s}_{\rm{P}})$ with a resulting only-quadratic
expression (2-local). The contraction techniques \cite{Boros2002} used
to reduce the locality incur an overhead of variables by introducing
ancillas (for a tutorial of a specific practical example see
Ref.~\cite{perdomo08}). This is not desirable because it increases the
search space from $2^{N_{\rm{P}}}$ to $2^{N_{\rm{Q}}}$, with
$N_{\rm{Q}}$ the number of variables $\mathbf{s}_{\rm{Q}}$ in the
resulting new quadratic expressions $H_{\rm{Q}}(\mathbf{s}_{\rm{Q}})$ as
the new representation from $H_{\rm{P}}(\mathbf{s}_{\rm{P}})$.
$\mathbf{s}_{\rm{Q}}$ is now the union of the health variables
$\mathbf{f}$, the wires $\mathbf{x}$, and the ancilla set $\mathbf{a}$.
The overhead is linear in our case as shown in
Fig.~\ref{f:qubit_resources}. The next challenge presented in
Fig.~\ref{f:ccfd-benchmark} towards implementing a real-world
application is that most likely there will be a quadratic term in
$H_{\rm{Q}}(\mathbf{s}_{\rm{Q}})$ representing qubit-qubit interactions
not present in the physical hardware. This will be the case unless one
specifically designs the layout of the quantum annealer hardware to match
the resulting connectivity graph dictated directly by the application
through $H_{\rm{Q}}(\mathbf{s}_{\rm{Q}})$. Representing the logical
graph within another graph is called the minor-embedding problem
\cite{Choi2011}. For the case of the connectivity graph predefined in
the D-Wave devices, also known as the \textit{chimera} graph, we use the
heuristic solver developed in Ref.~\cite{Cai-14}. As can be seen in
Fig.~\ref{f:qubit_resources}, the overhead is linear given the
relatively sparsity of the graphs resulting from the multiplier
circuits. This is an encouraging result given that the overhead for an
all-to-all connectivity graph embedded onto the chimera architecture is
quadratic in the number of variables.  A much larger problem than minor
embedding when embedding an application onto a limited connectivity
hardware graph is parameter setting.  For example, there is no rule of
thumb as to how strong the couplers for a set of
ferromagnetically coupled qubits defining a physical qubit should be. A
sweet-spot value is expected, however it is not easy to determine or
predict in the most general setting. In this work, and for all the
experiments on the DW2X, we use the strategy proposed in
Ref.~\cite{PerdomoOrtiz_arXiv2015a} for both setting the strength of the
ferromagnetic couplers and for the selection of gauges.  The final
challenge when embedding applications is the requirement that the
pseudo-Boolean function to be minimized has a low precision requirement
because analog QA machines operate on a limited precision
dictated by the intrinsic noise and finite dynamical range of parameters
found in these devices.

Summarizing, from our experience with applications, the CCFD instances
considered here are the best candidate to match each one of the
aforementioned requirements. The mapping from propositional logic to
PUBO is compact and efficient given that in the digital circuits
considered here all the input, outputs, health variables and wires are
all binary variables, the resulting QUBO graph is sparse enough that the
overhead to embed onto hardware is linear, and the randomly generated
instances have a higher intrinsic hardness compared to other random
spin glass previously studied, as shown in
Sec.~\ref{ss:hardness}.

Although we do not expect the intrinsic exponential scaling of this problem to disappear for the worst-case scenario by a mere change of representation or the solver used, the results could be different for each setting when computational times for typical instances are considered, and for the accesible problem sizes. The details and scaling slopes obtained for each of the approaches considered here are of extreme importance from a practical point of view, and used for addressing any meaningful advantage in the following sections.


\subsection{Hardness compared to other random spin-glass benchmarks}
\label{ss:hardness}

\begin{figure}
\includegraphics[width=\columnwidth]{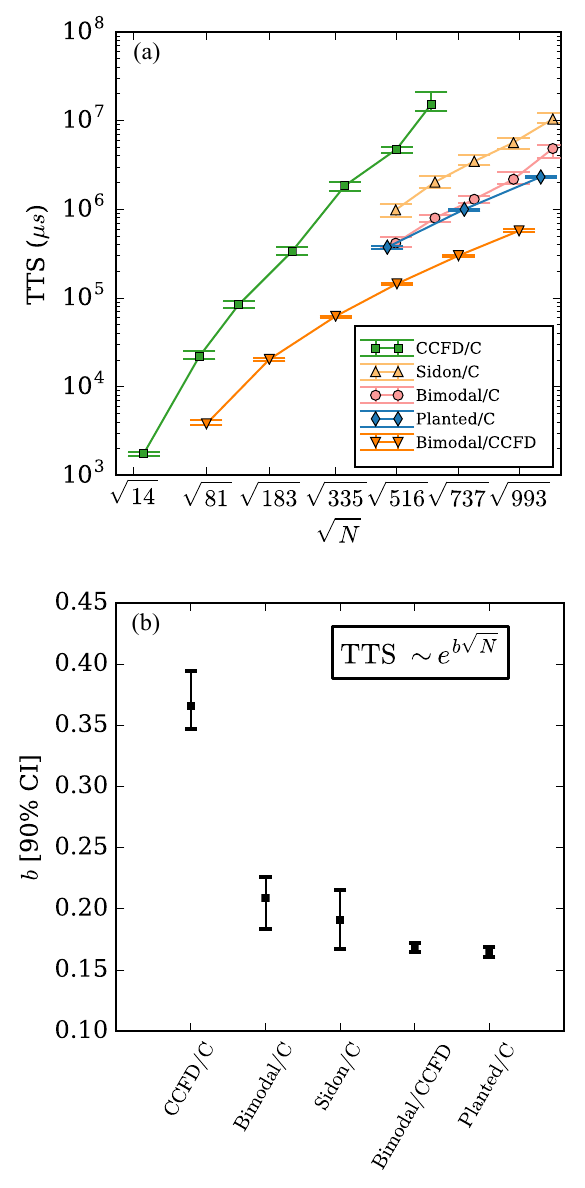}
\caption{
Hardness comparison of the chimera representation of the CCFD instances
with other representative random spin-glass problems from the
literature. (a) For consistency, all the time to solutions (${\rm
TTS}$) in $\mu$s are obtained with PTICM with the same single-core
processors.  (b) Note the steeper scaling [larger value for $b$ from a
fit of the ${\rm TTS}$ to ${\rm TTS} \sim \exp(b \sqrt{N})$] for the
CCFD problems compared to the other classes of random spin-glass
benchmarks. Data points correspond to the median values extracted from a
bootstrapping statistical analysis from $100$ instances per problem
size, with error bars indicating the $90$\% confidence intervals (CIs).
The different instance classes are described in the main text.}
\label{f:onlyPTICM}
\end{figure}

Figure \ref{f:onlyPTICM} addresses the hardness of instances embedded in
the chimera topology (C) generated from the CCFD data set by comparing
to random spin-glass problems used to benchmark the performance of
D-Wave quantum annealers [see Eq.~\eqref{eq:QUBO-C} for the actual
Hamiltonian to be minimized].  Bimodal instances were the first to be
used in benchmarking studies \cite{boixo:14,ronnow:14a} and are the
simplest to generate. For these, the available couplers in the D-Wave 2X
are randomly chosen to be $J_{ij} \in \{\pm 1$\} with biases $h_i  = 0$.
The reason why random bimodal instances are too easy for quantum and
classical algorithms alike is their high degeneracy resulting in a large
number of floppy spins. To overcome this problem,
Refs.~\cite{katzgraber:15,zhu:16} introduced couplers distributed
according to Sidon sets \cite{sidon:32} combined with postselection
procedures. These naturally increase the hardness of problems by
reducing degeneracy to a minimum and removing floppy spins. For the case
of Sidon instances \cite{zhu:16} the values of the couplers $J_{ij}$ are
randomly selected from the set $\{\pm 5, \pm 6, \pm 7 \}$, with $h_i  =
0$. Planted/C instances correspond to an attempt to increase the
hardness of random spin-glass instances (see Ref.~\cite{hen:15a}), but
with a known solution. For the data shown, we asked the main author in
Ref.~\cite{hen:15a}, if he could provide us with the hardest set of
instances he could generate; the only restriction being that they would
need to be generated randomly and not being postselected for hardness
as the rest of all the other families of instances here. The attempt
consisted of drawing the couplers from a continuous distribution instead
of from a discrete distribution as the one in the original paper,
Ref.~\cite{hen:15a}, or as in the case of the Bimodal and Sidon set
considered here.

Figure \ref{f:onlyPTICM}(a) illustrates that already for approximately
$600$ variables the CCFD/C instances are at least 1 order of magnitude
harder than Sidon/C which is the hardest set among the random spin-glass
problems. Figure \ref{f:onlyPTICM}(b) summarizes the asymptotic scaling
of each of these problem types, clearly separating our CCFD instances
from any of the random spin-glass instances, with Sidon and Bimodal
having roughly the same scaling. Here we assume that the
${\rm TTS}$ in $\mu$s can be fit to ${\rm TTS} \sim
\exp(b \sqrt{N})$] with $N$ the number of variables.  This conclusion is
independent of the percentile considered as shown in
Fig.~\ref{f:supp_ICMonly_scaling} in Appendix.~\ref{s:supp_hardness}.
Our results also show that the attempt to make hard planted instances
did not provide any additional hardness compared to the other random
spin-glass problems, at least when they are evaluated with PTICM.
Therefore, the CCFD/C instances are not only harder in terms of
computational effort, according to  ${\rm TTS}$, but also from a scaling
perspective.

The data set Bimodal/CCFD provides insights as to why these instances
are hard. There are three options of why these instances are
intrinsically harder than any other random data set explored here.  One
option is that the underlying CCFD graph defined by the QUBO problem for
each multiplier type has some sort of nontrivial long-range correlation
or a much higher dimensionality in such a way that the problems, when
minor embedded onto the chimera lattice, become harder than typical
chimera instances. Another explanation relies on the characteristic
value of biases $h$ and coupler values $J$ in the Hamiltonian
[Eq.~\eqref{eq:QUBO-C}] which could be responsible for the complex-to-traverse energy landscape. Furthermore, there could be
interplay between the two aforementioned options.  To address this
question, we generate Bimodal instances on the {\em native} QUBO graph
defined by multiplier circuits of varying sizes, denoted here as
``Bimodal/CCFD.'' If the underlying graph contains features that
intrinsically ``host'' hard instances, then one would expect that both
the scaling and ${\rm TTS}$ could be different than those on the chimera
graph.  Figure \ref{f:onlyPTICM} shows that the Bimodal/CCFD instances
happen to be even easier than the Bimodal instances embedded onto the
chimera graph.  This means that the intrinsic hardness of these CCFD
instances most likely is related to the structure and the relationship
between the specific biases $h$ and coupler $J$ values defining them.
Further studies are being performed to study this in more detail.

\subsection{Scaling analysis: application vs physics perspective}
\label{ss:physics-ccfd-scaling}

\begin{figure}[!h]
\includegraphics[width=\columnwidth]{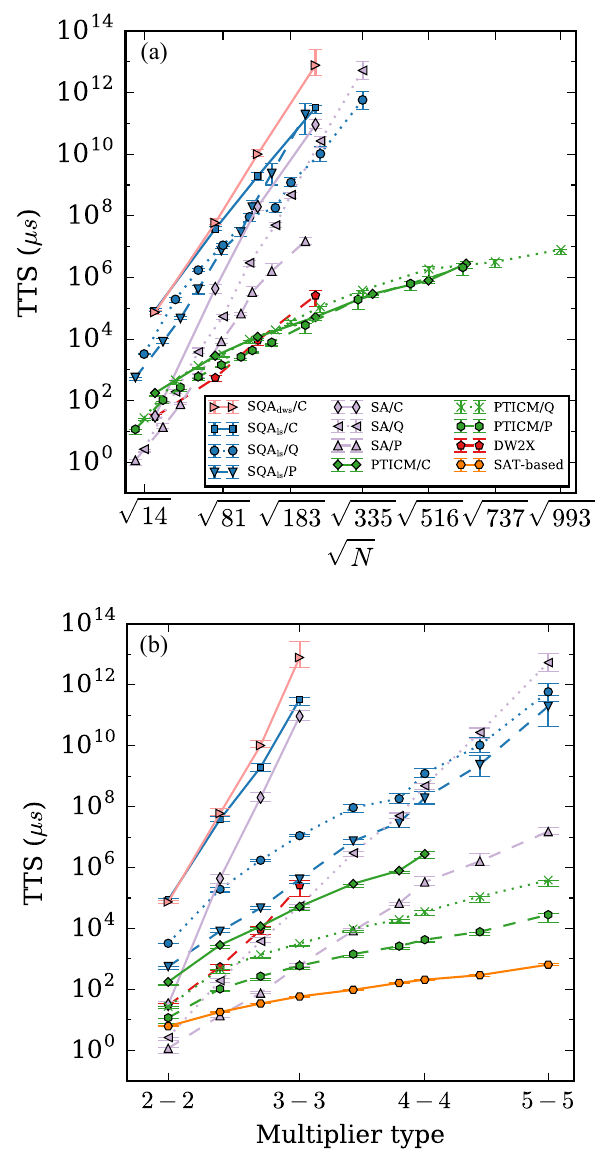}
\caption{Scaling analysis from (a) physics and (b) application-centric perspectives. The ${\rm TTS}$ is plotted as a
function of $\sqrt{N}$, with $N$ the number of spin variables in each of
the problem representations [PUBO (P), QUBO (Q), or chimera (C)]. Panel
(b) corresponds to $\sqrt{N_{\rm{gates}}}$, with $N_{\rm{gates}}$ the
number of gates regardless if we are considering symmetric multipliers,
mult[$n$-$n$] as in Fig.~\ref{f:ccfd_multk-k}, or asymmetric ones
(mult[$n$-$m$]).  The legend for the data sets depicted in panel (a) is
shared with panel (b), with SAT-based results only appearing in panel (b). SQA/C runs are performed with an optimized
linear schedule, as well as the DW2X schedule, marked with ``ls'' and
``dws'' subscripts, respectively, (details in
Appendix~\ref{s:dw2x_vs_qmc}).}
\label{f:physics-ccfd-scaling}
\end{figure}

Fig.~\ref{f:physics-ccfd-scaling} provides insights about the CCFD
instances from physics and application perspective.  While in
the former we analyze the scaling of computational resources via the ${\rm TTS}$ using the number of variables $\sqrt{N}$,
in the latter we analyze the resource requirements by the
application-specific variables, namely the type of multiplier used.  The
physics perspective here aims to answer questions about the performance
of QA compared to other classical solvers on a comparable
footing, ignoring for a moment that the instances are generated from a
specific application. For example, we compare here the performance of
QA to other classical and alternative quantum solvers on
instances represented on a chimera graph (C); similar to previous
extensive benchmarking work on synthetic random spin-glass instances. We
go beyond such studies and provide as well insights on the performance
of QA for the QUBO (Q) instances on their native graph
dictated by the CCFD application and also on hypothetical quantum
annealer devices capable of natively encoding up to quartic interactions
(P).

\begin{figure}
\includegraphics[width=\columnwidth]{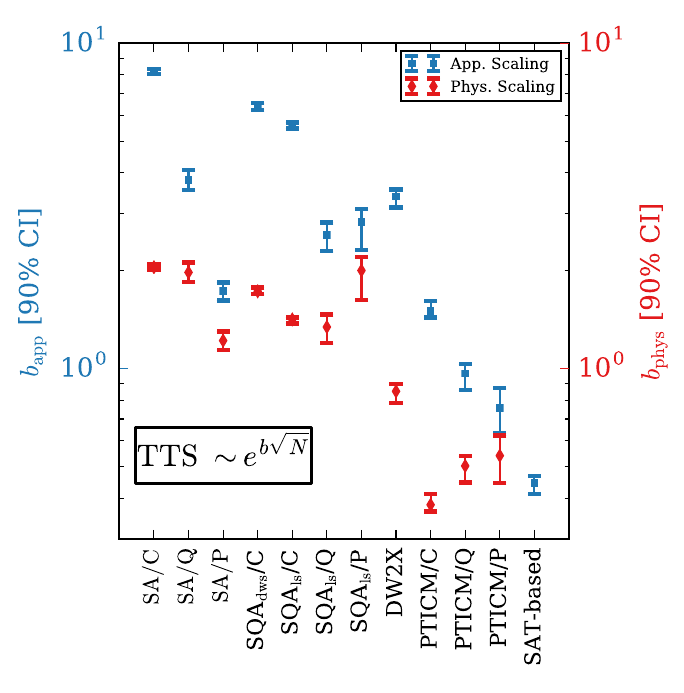}
\caption{Asymptotic scaling analysis.  The asymptotic scaling exponent
$b_{\rm app}$ refers to the multiplier representation, whereas $b_{\rm
phys}$ refers to the physical representation of the problem.  Data
points correspond to the median values extracted from a bootstrapping
statistical analysis from 100 instances per problem size, with error
bars indicating the 90\% CIs.}
\label{f:physics-ccfd-scaling_c}
\end{figure}

For the physics-scaling analysis we chose to plot the ${\rm TTS}$
computational effort as a function of $\sqrt{N}$, with $N$ being the
problem size in terms of number of spins, regardless of whether the
problem to be minimized is in a PUBO (P), QUBO(Q), or chimera (C)
format. This selection is motivated by the linear relation between any
pair of problem sizes $N_{\rm{P}}$, $N_{\rm{Q}}$ or $N_{\rm{C}}$ (see
Fig.~\ref{f:qubit_resources}) and the fact that the scaling for problems
on the quasiplanar chimera graphs is expected to be a stretched
exponential, largely due to its tree width approximately $\sqrt{N_{\rm{C}}}$, in
contrast to a tree width approximately $N$ characteristic of fully connected
graphs~\cite{Selby2014}.

The analysis from the application perspective aims for insights on the
performance where the sole purpose is to find the solution to the CCFD
problem. Here, it is natural to plot the ${\rm TTS}$ computational
effort as a function of a characteristic property of the circuit scaling
with the problem size, regardless if one considers a symmetric
multiplier, mult[$n$-$n$] or an asymmetric one, i.e., mult[$n$-$m$]. We
choose this quantity to be the number of gates in the circuit,
$N_{\rm{gates}}$ (or more precisely $\sqrt{N_{\rm{gates}}}$), which is
justified given the linear relationship between $N_{\rm{gates}} \propto
N_{\rm{P}} \propto N_{\rm{C}}$ illustrated in
Fig.~\ref{f:qubit_resources}, and the expected stretched exponential
behavior in $\sqrt{N_{\rm{C}}}$ discussed above for chimera graphs.

\paragraph{Limited quantum speedup} --- Figure
\ref{f:physics-ccfd-scaling}(a) compares the single-core computational
effort of SA, PTICM, SQA (with both linear and DW2X schedules), and the
experimental results obtained with the DW2X quantum annealer.
Represented with diamond symbols in Fig.~\ref{f:physics-ccfd-scaling_c}
and with values on the right axis, we plot the asymptotic analysis
performed by considering only the four largest sizes from each of the
data sets. From this scaling analysis and the value of the main scaling
exponent $b$ (slopes of curves in Fig.~\ref{f:physics-ccfd-scaling}) for
the chimera instances SA/C and DW2X, it can be seen that we also find
here limited quantum speedup (without optimizing annealing schedules)
\cite{mandra:16b} as found for the benchmarks on synthetic instances
used in the study by the Google Inc.~\cite{denchev:16}. From this
physics perspective, there seems to be even a quantum advantage when
comparing with SA at the PUBO level, SA/P, which happen to have a better
scaling than both of their quadratic counterparts, SA/Q and SA/C. The
values are close enough that one has to be careful because the real
$b_{\rm{phys}}$ (DW2X) might be larger than the calculated in our
analysis due to suboptimal annealing time
\cite{ronnow:14a,Albash2017,mandra:16b}. On the other hand, note that
the quantum advantage at the level of the same representation where
$b_{\rm{phys}}$(DW2X)~$\ll b_{\rm{phys}}$(SA/C) also holds against any
of the optimized SQA/C implementations, either with a linear or DW2X
schedule. Although we believe that it is very unlikely that suboptimal
time can change our \textit{limited quantum speedup} conclusion because
$b_{\rm{phys}}$(DW2X)~$\ll b_{\rm{phys}}$(SA/C), we note that optimized
SQA corroborates these claims.  This scaling advantage already yields a
difference of approximately 6 orders of magnitude on a single-core CPU
in the ${\rm TTS}$ between DW2X and SA/C for the largest problem studied
(mult[4-4]). It is important to remind the reader that our results are with a fixed annealing time, and although we justify that it would be very unlikely that the slope of DW2X could match that of SA/C, the best practice to have conclusive limited quantum speedups results would be by optimizing the annealing time in the quantum-annealer runs~\cite{ronnow:14a,Job2018}. Exploration of the impact of the optimal annealing time in the CCFD instances could be an interesting piece of work in its own and it is left as future work.

\paragraph{SQA vs DW2X and impact of the annealing schedule} --- From a
computational prefactor perspective note that the computational effort
for the DW2X is smaller by 6 to 8 orders of magnitude than the
SQA/C implementations with either linear or the D-wave schedule. It is
important to note that the ${\rm TTS}$ in
Figs.~\ref{f:physics-ccfd-scaling}(a) and
\ref{f:physics-ccfd-scaling}(b) is for SQA as a classical computational
solver. For a fair comparison of the scaling of SQA to that of a
physical quantum annealer such as the DW2X, the SQA $\rm TTS$ results
must be divided by $N$ to account for the intrinsic parallelism in
QA, as illustrated in Fig.~\ref{f:dw2x_qmc}. Further
analysis of the scaling comparison of SQA and the DW2X device can be
found in Appendix~\ref{s:dw2x_vs_qmc}.

Figures \ref{f:physics-ccfd-scaling}(a)
and \ref{f:physics-ccfd-scaling}(b) illustrate that the
selection of a poor schedule (the D-wave schedule in this case in
comparison to the simpler linear one) can have a significant impact in the
computational efficiency of SQA as the classical computational solver. As
discussed in Appendix~\ref{s:dw2x_vs_qmc}, most likely the difference is
only at the level of a prefactor and most likely it is not a scaling
advantage. Whether there are schedules that can change the asymptotic
scaling remains an open question. In Appendix~\ref{s:dw2x_vs_qmc} we also
discuss that although there seems to be a scaling
advantage of the DW2X over the SQA simulations, the results are also
inconclusive given that the scaling of the DW2X might be slightly
different due to any  suboptimal annealing times. We leave it to future
work to optimize the annealing time of the DW2X because it is beyond the
scope of this work given the sizable computational requirements needed.

\paragraph{QA performance for Hamiltonians with
higher-order interactions} --- A question not addressed to date is the
performance comparison between QA architectures with
2-local and $k$-local ($k > 2$) interactions within the scope of
real-world applications.  For example, the CCFD mapping used in this
work (see Appendix~\ref{s:mapping} for details) natively contains cubic
(3-local) and quartic (4-local) interactions and one might think a
quantum annealer natively encoding those might have an advantage over
2-local terms.  Perhaps one of the most remarkable findings in this
study from our SQA simulations is that working directly with a
Hamiltonian containing such quartic interactions does not seem to help
QA with a transverse field, because
$b_{\rm{phys}}$(SQA/Q)~$< b_{\rm{phys}}$(SQA/P).  Note that this result
is in contrast to the behavior of the classical algorithms considered
here. In the case of SA there seems to be an advantage for solving the
instances in the PUBO representation over the QUBO
[$b_{\rm{phys}}$(SA/Q)~$ > b_{\rm{phys}}$(SA/P)]. In the case of PTICM,
$b_{\rm{phys}}$(PTICM/Q)~$\approx b_{\rm{phys}}$(PTICM/P).  These
remarks on the physics scaling have a significant impact on the scaling
from the application perspective. Note that while in all the classical
methods there is a clear preference to solve the problem in the PUBO
representation, the case of SQA shows no advantage for the quantum
annealer in the PUBO representation. In contrast, as shown in
Fig.~\ref{f:supp_natural_scaling} for the higher percentiles above the
median, there seems to be a slight preference of SQA/Q over SQA/P not
only in the absolute value of computational effort measured in $\rm TTS$
[see the last data point in Fig.~\ref{f:supp_natural_scaling}(d)], but also
in scaling terms.

The insight to be extracted from the SQA simulations in the context of
this CCFD application is that simply adding higher-order terms would not
necessarily imply any enhancement in the performance. This result is
striking for two reasons. 

First, because in the application scaling we plot the $\rm TTS$ results
vs $\sqrt{N_{\rm{gates}}}$, then when changing representations from PUBO
to QUBO, there is a natural tendency for $b_{\rm{app}}$(Q)$/b_{\rm{app}}$(P)~$> 
b_{\rm{phys}}$(Q)$/b_{\rm{phys}}$(P). This is because $N_{\rm{Q}}$ is
always greater than $N_{\rm{P}}$, and therefore even in the case of
comparable physics scaling slopes as is the case of PTICM with
$b_{\rm{phys}}$(PTICM/Q) $ \approx b_{\rm{phys}}$(PTICM/P) this would
imply that
\begin{equation}
\!\!\!\!\!\!\!\!\!\!\!\!\!\!\!\!\!{\rm TTS}_{\rm{PTICM/P}} 
\sim e^{b_{\rm{phys}} \sqrt{N_{ \rm{P} }}} 
\sim e^{b_{\rm{phys}} \sqrt{\alpha_{\rm{P} 
\leftarrow \rm{g}}}\sqrt{N_{ \rm{gates} }}} ,\nonumber
\end{equation}
while
\begin{equation}
{\rm TTS}_{\rm{PTICM/Q}} 
\sim e^{b_{\rm{phys}} \sqrt{N_{ \rm{Q} }}} 
\sim e^{b_{\rm{phys}} \sqrt{\alpha_{\rm{Q} 
\leftarrow \rm{P}} \alpha_{\rm{P}
\leftarrow \rm{g}}}\sqrt{N_{ \rm{gates} }}} ,\nonumber 
\end{equation}
valid in the asymptotic limit $N_{ \rm{gates}} \gg 1$ of interest here.
Therefore,
\begin{equation}
b^{\rm{Q}}_{\rm{app}} = 
b_{\rm{phys}} \sqrt{\alpha_{\rm{Q} 
\leftarrow \rm{P}} \alpha_{\rm{P} 
\leftarrow \rm{g}}} >  b^{\rm{P}}_{\rm{app}} = 
b_{\rm{phys}} \sqrt{\alpha_{\rm{P} 
\leftarrow \rm{g}}}. \nonumber
\end{equation}
Here we use that, in this limit, $N_{\rm{P}} \sim \alpha_{\rm{P}
\leftarrow \rm{g}} N_{\rm{gates}}$ and $N_{\rm{Q}} \sim \alpha_{\rm{Q}
\leftarrow \rm{P}} N_{\rm{P}}$, and  both, $\alpha_{\rm{Q} \leftarrow
\rm{P}}$ and $\alpha_{\rm{P} \leftarrow \rm{g}}$ are greater than $1$,
as shown in Fig.~\ref{f:qubit_resources}.

Second, the penalties of the locality reduction ancillas change the
energy scale and it is expected that stochastic solvers such as SA
(which heavily depend on the barriers in the energy landscape) also
suffer from the new QUBO energy landscape with taller barriers. This is
indeed what we observe because $b_{\rm{phys}}$(SA/Q)~$>
b_{\rm{phys}}$(SA/P). Note that PTICM seems to be more resilient to
these barriers and, as discussed before,
$b_{\rm{phys}}$(PTICM/Q)~$\approx b_{\rm{phys}}$(PTICM/P). Both of these
driving forces would imply that the application perspective scaling of
QA working in the PUBO representation should be better
than in the QUBO representation and it is not what we observe here. This
second explanation is reasonable and a good indication that SQA is doing
a good job at not ``feeling'' these taller barriers, something that
could be explained by means of quantum tunneling.

From the first argument it follows that $b_{\rm{app}}$(SQA/Q)~$\approx
b_{\rm{app}}$(SQA/P), implying that $b_{\rm{phys}}$(SQA/Q) $<
b_{\rm{phys}}$(SQA/P) which is quite distinctive and different from what
we observe in the classical approaches. It is clear that SQA is having a
harder time traversing the PUBO energy landscape and finding the ground
state in this representation, despite the smaller problem size.  One
plausible explanation is that the transverse-field implementation is not
powerful enough to take advantage of the compactness of the PUBO energy
landscape. We thus emphasize that any development of new architectures
with $k$-local couplers with $k > 2$ should be accompanied by
other developments, that could enhance its computational power, such as the inclusion of more sophisticated driver Hamiltonians. 

\paragraph{Impact of the limited connectivity} --- Here we address the
issues that occur with limited-connectivity hardware (see
Fig.~\ref{f:ccfd-benchmark}). From the physics scaling perspective,
Fig.~\ref{f:physics-ccfd-scaling}(a) and
Fig.~\ref{f:physics-ccfd-scaling_c} show that there are no major effects
in solving the problems with the QUBO or with the chimera
representation. This seems to be a common feature across classical and
quantum approaches.  Following the argument just previously made in the
case of the PUBO vs QUBO discussion, we show that $b_{\rm{phys}}(\rm{Q})
\approx b_{\rm{phys}}(\rm{C})$ and $N_{\rm{C}} \sim \alpha_{\rm{C}
\leftarrow \rm{Q}} N_{ \rm{Q} }$, implies that  $b_{\rm{app}}(\rm{C}) >
b_{\rm{app}}(\rm{Q})$. Here, $\alpha_{\rm{C} \leftarrow \rm{Q}} =
3.5026$ from Fig.~\ref{f:qubit_resources}. Although it has always been
expected that more connectivity should be better, having a quantum
annealing device with more connectivity can have a significant impact
when solving real-world applications. Our results, within the context of
the CCFD application, show that the advantage here is not simply an
overall prefactor improvement in the $\rm TTS$ but that an important
asymptotic scaling advantage is expected as well. As a reminder to the reader, this \textit{in-silico} advantage from SQA will be matched by a quantum hardware implementation only under the assumption that the asymptotic scaling of SQA "mimics" the performance of QA. As stated in the introduction, this is an unsettled question and beyond the scope of our work.

\paragraph{Comparison of QA with generic and tailored
algorithms for CCFD} --- The main question that motivated this study was
if QA can efficiently solve CCFD problems.  Figures
\ref{f:physics-ccfd-scaling}(b) and \ref{f:physics-ccfd-scaling_c} show
that from the application perspective the scaling of the DW2X quantum
annealer and of any of the SQA variants considered here does not look
favorable for QA.  In fact, the DW2X does not even scale
better than simulated annealing (SA/P).  One of the major challenges for
devices with a small finite graph degree connectivity (such as the DW2X
with the chimera topology) is that to solve real-world applications it
carries all the qubit overhead from the transformations PUBO to QUBO
over to chimera.  However, the application scaling can be improved.  For
example, reducing the number of qubits needed to represent the
application would most likely improve the scaling performance. In
Sec.~\ref{s:mapping} we present an alternative and more efficient
mapping that we aim to explore in further studies. It is important to
note that the new mapping improves the performance of SA and PTICM
accordingly. Note that from all the approaches, the most efficient with
the best scaling is a SAT-based solver developed by our team for this
study which excels in this strong-fault model-based diagnosis of
multiplier circuits. The SAT-based solver does not depend on any of the
PUBO, QUBO, or chimera representation because it has the advantage of
constructing its own variable representation and set of satisfiability
constrains directly from the propositional logic level shown in Fig.~\ref{f:ccfd-benchmark}(a). Although all other classical
stochastic solvers used (SA, PTICM, and SQA) might work directly with
the propositional logic as well, the evaluation of the cost function
would be highly nonlocal compared to the evaluation of the difference in
energy required for the Metropolis update in the case of the polynomial
evaluation. By nonlocal we mean that if we were to work with only the
fault variables and use the propositional logic instead of constructing
the PUBO and including the internal wire variables, then we would need
to propagate from inputs all the way through each gates and their health
status assignment to obtain the predicted outputs and subsequently an
effective energy that can be used in the Metropolis update. For every
pair considered in an Metropolis update, the whole process needs to be
applied and later subtracted. In contrast, in any of the polynomial
representations, because all the fault and wire variables are
considered, the evaluation of the energy difference can be applied very
efficiently by only considering the few terms that change the energy by
the respective variable flip. Most importantly, because the main point
of this contribution is to compare with algorithms that could be
implemented in QA architectures, we did not explore this implementation.
We leave it as an open question whether algorithms like SA can have a
better scaling on that propositional logic representation.

\section{Conclusions}
\label{s:conclusions}

Regardless of the substantial efforts in benchmarking, the study of
early generation quantum annealers has been done exclusively with
synthetic spin-glass benchmarks. However, comprehensive studies
comparing several quantum and classical algorithmic approaches,
including state-of-the-art tailored solvers for real-world applications,
had been missing. In this work we present a comprehensive benchmarking
study on a concrete application, namely the diagnosis of faults in
digital circuits, referred in the main text as CCFD. More specifically,
we provide insights on the performance of QA in the
context of the CCFD instances by performing an asymptotic scaling
analysis involving five different approaches: QA
experiments on the DW2X compared to three classical (SA, PTICM, and a
CCFD-tailored SAT-based solver), and extensive QMC simulations, most of
them on three different problem representations (PUBO, QUBO on the
native CCFD graphs, and QUBO on the DW2X chimera topology), for
instances of multiplier circuits of varying size. It is important to
note that by asymptotic analysis we refer to conclusions drawn from the
largest problem sizes accessible to us to experiment with in each of
these approaches.

We have analyzed the problem with two foci: a {\it physics perspective}
and an {\it application-centric perspective}. The emphasis of the
physics perspective is similar to previous representative benchmark
studies
\cite{johnson:11,dickson:13,boixo:13a,katzgraber:14,boixo:14,ronnow:14a,katzgraber:15,heim:15,hen:15a,rieffel:15,boixo:16,denchev:16,king:19,Albash2017},
which aim at probing the computational resources of QA,
and to answer questions such as whether it is even possible in synthetic
data sets to prove an asymptotic quantum speedup, or to address the role
of quantum tunneling, among other open questions in the field. Within
our physics perspective we add several issues not thoroughly considered
in other benchmark studies. For example, what is the impact in the
computational scaling of solving the problem directly with Hamiltonians
natively encoding many-body interactions beyond pairwise as those
naturally appearing in real-world applications? What is the impact in
the scaling from solving the problem instances on (hypothetical)
physical hardware with different qubit connectivity constrains, e.g., by
comparing the QA performance on connectivity graphs
dictated by the CCFD instances and the minor-embedded representation in
the DW2X chimera graph?

From this physics perspective we show that our instances are hardest
when compared to any of the proposed random spin-glass instances (see
Fig..~\ref{f:onlyPTICM}). Intrinsic hardness is one of the long
sought-after features when performing benchmark studies
\cite{katzgraber:14,hen:15a,king:19}, therefore making our CCFD
instances currently the best candidate for benchmarking the next
generation of quantum annealers. In particular, because these problems
stem from real-world applications, in contrast to random synthetic benchmarks on the native D-wave's chimera graph which have been dulled not only for giving an advantage to the hardware but also for lacking practical importance~\cite{AaronsonBlogGooglePaper,AaronsonBlogDWbenchmarks}.

We also address the question of whether SQA can reproduce the scaling of
the DW2X for the CCFD application. Although the results in
Fig.~\ref{f:dw2x_qmc} might lead to the conclusion that clearly SQA has
a different scaling than the DW2X, the fact that most likely the DW2X is
running at a suboptimal annealing time might be distorting the scaling
and resulting in a better apparent scaling.  More extensive studies with
enough data points --- where one can optimize for the optimal annealing
time --- might reveal the real scaling of the device. Although this is,
in principle, feasible on quantum annealers, the main challenge might
rely on SQA simulations which are already at the limit of what is
computationally feasible.  In Appendix~\ref{s:dw2x_vs_qmc} we discuss
the apparent different scaling of SQA with a linear schedule compared to
SQA with the same schedule as the D-Wave device and the challenges on
drawing any meaningful conclusions about the difference in scaling
between the DW2X and SQA. From the choice of schedule perspective, we
find that within SQA as a solver, the linear schedule seems to be more
efficient, but most likely not bringing any scaling advantage.

When compared on the same representation (either native-QUBO or chimera-QUBO) we show that both, SQA and the DW2X have a limited quantum
speedup by showing a scaling advantage over SA. We arrive at this
conclusion assuming the DW2X scaling obtained here is not drastically
affected by the nonoptimal annealing time, which is very unlikely due
to the large difference in the slopes between SA and SQA and DW2X.
These results confirm the presence of quantum tunneling in the DW2X; a
quantum speedup restricted to sequential algorithms~\cite{mandra:16b}
similar to the Google Inc.~study on the weak-strong clusters
instances~\cite{denchev:16}.  One important highlight here is that ours
is an alternative demonstration on instances generated from a concrete
real-world application and where the multispin co-tunneling needs to
happen more often on the strongly ferromagnetically coupled physical
qubits encoding the logical units from the original QUBO problem.
Although it is encouraging to see that such co-tunneling events seem to
be happening in the hardware at the problem sizes considered here, the
minor-embedding mapping logical variables into physical qubits in the
hardware usually involves the generation of long ``chains.'' Further
studies need to be performed using larger instances to see if this
advantage remains, and where longer ``chains'' with ten or more qubits
would be more frequent. The comparison against other generic solvers
like PTICM or tailored solvers like the SAT-based solver developed here,
were not favorable for our SQA simulations and DW2X experiments. It is
important to consider that both, SQA simulations and DW2X experiments,
were done with stoquastic Hamiltonians as the only ones available in
current hardware. It is expected that nonstoquastic Hamiltonians will
bring a boost in performance \cite{Hormozi2016,Nishimori2017}, although
it is an open question if they will have any asymptotic scaling
advantage.

The application-centric perspective is more challenging and raises the
bar significantly for quantum annealers. Here, we find that the tailored
SAT-based algorithm performs best. We note that the performance is even
better than the PTICM algorithm, which is currently the state of the art
in the field \cite{biere16splatz}. Although the results using quantum optimization approaches as seen from
the application perspective are not that encouraging, our study suggests
next steps to be taken in the field of quantum optimization. First,
there is a clear need for higher-connectivity devices. Second, our SQA
results suggest that adding higher-order qubit interactions
\cite{Chancellor2017,Strand2017} to new hardware might require also the
addition of more complex driving Hamiltonians.  

This rather extensive study should be considered as a baseline for future application studies. We do emphasize, however, that the conclusions should be interpreted within the context of the particular CCFD application. Furthermore, the results are for the specific case of conventional QA with a transverse-field driver. The poor performance of QA should be seen as an incentive for the community to address important missing ingredients in the search for quantum advantage for real-world applications.  Other variable efficient mappings (as shown in Appendix~\ref{ss:implicit}) could also provide a performance boost. We note that the latter should also provide an advantage for classical solvers, because larger systems could be studied. A detailed performance comparison of our CCFD benchmarks to other mapping strategies ~\cite{Bian2016} will be done in a subsequent study.

Further adding other
features, such as better control of the annealing schedules
via ``seeding'' of solutions~\cite{PerdomoSAQC2010}, and the subsequent developments of classical-quantum hybrid heuristic strategies~\cite{PerdomoSAQC2010,Chancellor2017a,Chancellor2017b,Karimi2017} will likely lead to
breakthroughs in quantum optimization.  However, more simulations are
needed to guide the design of new machines.

\begin{acknowledgments} 

The work of A.P.O. is supported in part by the AFRL Information
Directorate under Grant No. F4HBKC4162G001, the Office of the Director of
National Intelligence (ODNI), and the Intelligence Advanced Research
Projects Activity (IARPA), via IAA 145483.
Z.Z.~and H.G.K.~acknowledge support from the National Science Foundation
(Grant No.~DMR-1151387).  The work of H.G.K.~and Z.Z.~is supported in
part by the Office of the Director of National Intelligence (ODNI),
Intelligence Advanced Research Projects Activity (IARPA), via MIT
Lincoln Laboratory Air Force Contract No.~FA8721-05-C-0002.
The views and conclusions contained herein are those of the authors and
should not be interpreted as necessarily representing the official
policies or endorsements, either expressed or implied, of ODNI, IARPA,
AFRL, or the U.S. Government.  The U.S. Government is authorized to
reproduce and distribute reprints for Governmental purpose
notwithstanding any copyright annotation thereon.
We thank Delfina Garcia-Pintos for the implementation of the libraries
used in the bootstrapping analysis of the slopes and their confidence
intervals of our asymptotic analysis. We also thank Tayo Oguntebi for
initial support and discussions related to this CCFD application. We also thank Catherine McGeoch for feedback on an earlier version of this manuscript. 

\end{acknowledgments}


\appendix 

\section{QA for combinatorial optimization problems}\label{s:qa}

\begin{figure}
\centering
\includegraphics[width=0.45\textwidth]{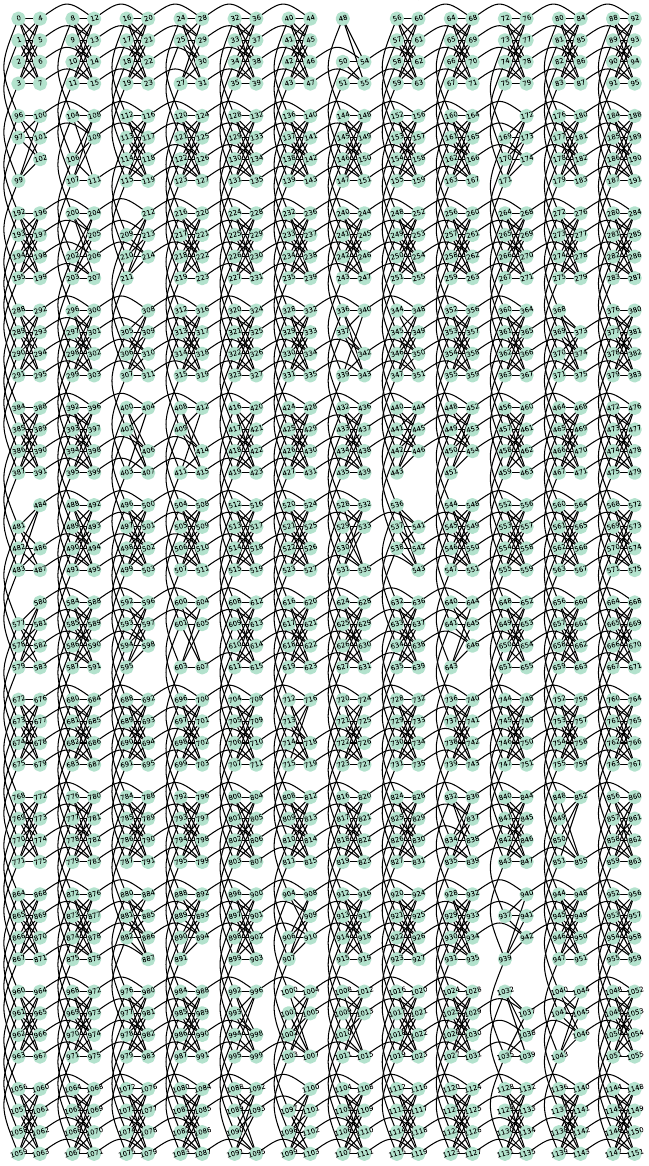}
\caption{
Device architecture and qubit connectivity. The array of superconducting
quantum bits is arranged in $12\times 12$ unit cells that consist of eight
quantum bits each. Within a unit cell, each of the four qubits on the
left-hand partition (LHP) connects to all four qubits on the right-hand
partition (RHP), and vice versa. A qubit in the LHP (RHP) also connects
to the corresponding qubit in the LHP (RHP) of the units cells above and
below (to the left and right of) it. Edges between qubits represent
couplers with programmable coupling strengths. We show only the 1097
functional qubits out of the 1152 qubit array.}\label{fig:chimera}
\end{figure}

The quantum hardware employed consists of $144$ unit cells with eight
qubits each, as characterized in Refs.~\cite{harris2010,johnson:11}.
Postfabrication characterization determined that only $1097$ qubits 
from the $1152$ qubit array can be reliably used for computation, as shown in
Fig.~\ref{fig:chimera}. The array of coupled superconducting flux
qubits is, effectively, an artificial Ising spin system with
programmable spin-spin couplings and magnetic fields. It is
designed to solve instances of the following
(NP-hard~\cite{Barahona1982}) classical optimization problem: given a
set of local longitudinal fields $\{h_i\}$ and an interaction matrix
$\{J_{ij}\}$, find an assignment $\mathbf{s^*} = s^*_1 s^*_2 \cdots
s^*_N$, that minimizes the objective function $E: {\{-1, +1\}}^N
\rightarrow \mathbb R$, where
\begin{equation}\label{eq:QUBO-C}
E(\mathbf{s}_{\mathrm{C}}) 
= \sum_{1 \le i \le N} h_{i} s_i  + \sum_{1 \le i<j\le N} J_{ij} s_{i} s_{j}.
\end{equation}
Here, $\abs{h_i} \le 2$, $\abs{J_{ij}} \le 1$, and $s_i \in \{+1,-1\}$.
The subscript ``C'' is to emphasize that the spins are within the
\textit{chimera} graph, and to differentiate these from the other two
representations studied in the paper at the PUBO
($\mathbf{s}_{\mathrm{P}}$) and QUBO ($\mathbf{s}_{\mathrm{Q}}$) level,
respectively.

Finding the optimal set of variables $\mathbf{s^*}$ is equivalent to
finding the ground state of the corresponding Ising classical
Hamiltonian,
\begin{equation}\label{h-ising}
H_{p}  =  \sum^N_{1 \le i \le N} h_{i}\sigma_{i}^{z}  + \sum^N_{1 \le i<j\le N} J_{ij}\sigma_{i}^{z} \sigma_{j}^{z},
\end{equation}
where $\sigma_{i}^{z}$ is a Pauli $z$ matrix acting on the $i$th spin.

Experimentally, the time-dependent quantum Hamiltonian implemented in
the superconducting-qubit array via
\begin{equation}\label{h-AQO}
H(\tau)  = A(\tau) H_b + B(\tau) H_p, \quad \quad \tau= t/t_{a},
\end{equation}
with $H_b  = - \sum_i \sigma^{x}_{i}$ the transverse-field driving
Hamiltonian responsible for quantum tunneling between the classical
states constituting the computational basis, which is also an eigenbasis
of $H_p$. The time-dependent functions $A(\tau)$ and $B(\tau)$ are such
that $A(0) \gg B(0)$ and $A(1) \ll B(1)$. In Fig.\ref{f:dw2x_qmc}(a), we
plot these functions as implemented in the experiment. $t_{a}$ denotes
the time elapsed between the preparation of the initial state and the
measurement, referred to hereafter as the \textit{annealing time}.

QA as an algorithmic strategy to solve classical optimization problems
exploits quantum fluctuations and the adiabatic theorem of quantum
mechanics. This theorem states that a quantum system initialized in the
ground state of a time-dependent Hamiltonian remains in the
instantaneous ground state if the Hamiltonian changes sufficiently
slow. Because the ground state of $H_p$ encodes the solution to the
optimization problem, the idea behind QA is to adiabatically prepare
this ground state by initializing the quantum system in the
easy-to-prepare ground state of $H_b$, which corresponds to a
superposition of all $2^N$ states of the computational basis, and then
slowly interpolating to the problem Hamiltonian, $H(\tau=1) \approx
H_p$.

In a realistic experimental implementation, the quantum processor will
operate at a finite temperature, and in addition to thermal
fluctuations, other types of noise are unavoidable, leading to
dissipation processes not captured in $H(t)$. Deviations from
adiabaticity affecting the performance of the quantum algorithm seem to
be a delicate balance between the quantum coherence effects and the
interaction with the environment, responsible for, e.g., thermal
excitation (relaxation) processes out of (into) the ground
state~\cite{AlbashNJP2012,PerdomoOrtiz2012_LPF}.

Determining the optimum value of $t_a$ is an important and nontrivial
problem in itself. To the best of our knowledge this question related to
the scaling of $t_a$ in a noisy environment is still largely unexplored,
with progress only in the case of canonical
models~\cite{Smelyanskiy_PRL2017}. From an experimental standpoint, the
main limitation is the limited size of the available quantum devices,
but now with new generation of devices with more than $2000$ qubits the
question is within reach in the case of synthetic data
sets~\cite{Albash2017}. Studying this question within the context of
real-world applications is now within reach. We leave this study
for future work.

\section{Methods}~\label{s:methods}

\paragraph{Simulated Quantum Annealing (SQA)} --- QMC simulations are
performed using a variant of the continuous time QMC
algorithm~\cite{rieger:kawashima}, which we refer to here and in the
main text as SQA. We build clusters in the imaginary time direction in
the same way as done in Ref.~\cite{rieger:kawashima}. However, because
here we study frustrated systems, we do not build clusters in the
spatial directions.  To flip segments of finite imaginary time extent,
we use the Metropolis algorithm \cite{interplay}. This algorithm was
also used in previous benchmark studies of the D-Wave
devices~\cite{boixo:14,ronnow:14a}.

\begin{figure}[!h]
\includegraphics[width=0.86\columnwidth]{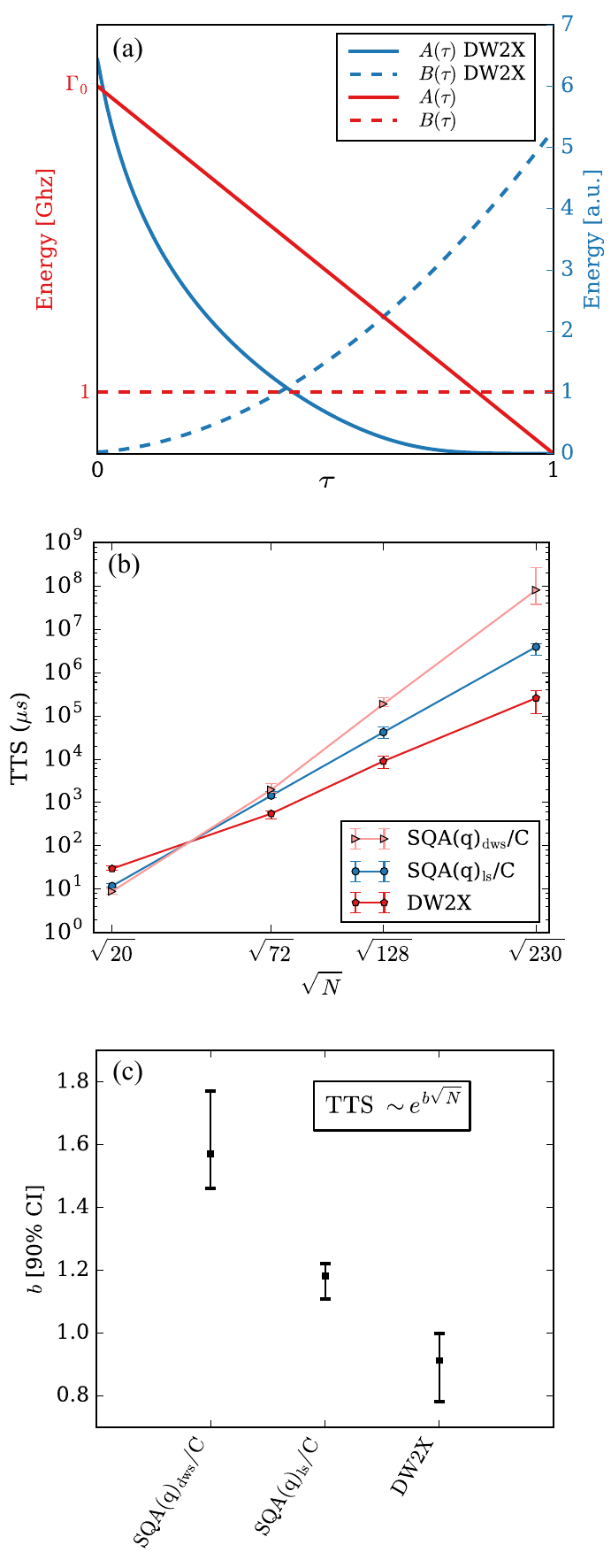}
\caption{
(a) Details for the different annealing schedules used in this work.
Panels (b) and (c) show a comparison of the DW2X experimental results
and SQA simulations of hypothetical QA devices with a
DW2X-like [SQA(q)$_{\rm{dws}}$] and with a linear annealing schedule
[SQA(q)$_{\rm{ls}}$]. Data points correspond to the median values
extracted from a bootstrapping analysis from $100$ instances per problem
size, with error bars indicating the 90\% CIs.}
\label{f:dw2x_qmc}
\end{figure}

In the SQA simulations we use a linear schedule. We fix the diagonal
interaction strength $B(\tau) = 1$ and vary the transverse-field
strength as $A(\tau) = \Gamma_0 (1 - \tau)$, where $\tau$ is the
annealing time and $\Gamma_0$ is the initial transverse-field strength;
see Fig.~\ref{f:dw2x_qmc}.  We use different values of $\Gamma_0$ for
different problem representations. $\Gamma_0 = 0.8$ for PUBO, $\Gamma_0
= 1.6$ for QUBO, and $\Gamma_0 = 6$ for instances on the chimera graph.
These representations are referred as ``P,'' ``Q,'' and ``C,''
respectively, in the main text. In addition, we also implement the
$A(\tau)$ and $B(\tau)$ annealing schedule used in the DW2X device, as
depicted in Fig.~\ref{f:dw2x_qmc}.

\paragraph{Estimation of the time-to-solution ($\rm TTS$)} --- For
stochastic algorithms, the time-to-solution depends on the desired
confidence, i.e., the probability $P$ required such that the solver
produces the target solution. For example, in all previous studies the
level of certainty required from the solver was 99\%, i.e., $P=0.99$,
and the relevant metric, denoted $R_{99}$ is the number of repetitions
needed such that the probability that the solution is found is at least
once is 99\%.  Let us denote by $p_{\rm s}$ the success probability to
obtain the target solution in a single execution or repetition of the
solver. Because the probability $F$ of {\em not} observing the solution
after $R_{P}$ repetitions is $F = (1-p_{\rm s})^{R_P} = 1-P$, the number
of repetitions $R_{99}$ needed to obtain the desired solution with
probability at least 99\% is
\be
R_{99} = \ceil{ \frac{\log(1-0.99)}{\log(1-p_{\rm s})}}.
\ee
Therefore, the time-to-solution ($\rm TTS$) under this criteria is the
product of $R_{99}$ times the time it takes to perform one execution or
each repetition, $t_{\rm{rep}}$:
\be 
\rm{TTS}  = t_{\rm{rep}} R_{99}.
\ee
For the DW2X, $t_{\rm rep}$ was set to the annealing time of 5 $\mu s$. For
SA, QMC, and PTICM, $t_{\rm rep}$ can be estimated as:
\be
t_{\rm rep} = t_{\rm{su}} N \rm{MCS}_{\rm{opt}},
\ee
with $\rm{MCS}_{opt}$ the optimal number of Monte Carlo sweeps (MCS),
i.e., the number of MCS that minimizes the $\rm TTS$ and $t_{\rm{su}}$ the time it takes to make a MC spin update. In the case of PTICM, the values of $\rm{MCS}_{opt}$ include a factor of 120 coming from the four replicas and 30 temperatures considered in our implementation. Additionally, we multiply by a factor of 1.2 to account an estimated 20\% overhead coming from other steps in the PT implementation and not present in SA, such as swaps of configurations and cluster updates.  Here we optimize
the $\rm TTS$ per instance to obtain the best scaling for each
algorithm.  The computational effort of the algorithm optimization of
this procedure compared to optimizing over different annealing times as
proposed in Ref.~\cite{ronnow:14a} should yield comparable scaling
results. We prefer this approach because it requires the same
computational effort as analyzing the data at different annealing times
and it provides more reliable information on the intrinsic difficulty of
each instances. For example, in the limit of very large annealing time,
where all the instances have probability 1, most instances have the
same computational effort and it is not possible to identify which
instances are intrinsically harder. This is related to the problem with
reported DW2X scaling for small instances, for which the minimum
available annealing time is greater than optimal.  Note that $p_{\rm s}$
is a function of the number of MCS, in the same way that it is a
function of the annealing time in the case of the DW2X. Therefore, we
estimate $p_{\rm s}$ for different values for the number of MCS,
calculate $R_{99}$ and from the values considered we select the optimum.
Since one MCS involves $N$ updates~\cite{katzgraber:09e}, with $N$ the
total number of spins in the problem, then to calculate the
computational effort we need to multiply by $N$ and by the effective
time it takes to perform and evaluate each of these updates. The value
$t_{\rm su}$ is different for each of the algorithms (e.g., SA vs SQA)
and for each of the different representations (QUBO, PUBO, or chimera).
The times estimated and used for the case of the CCFD instances are:
$t^{\rm{SA/P}}_{\rm{su}} = t^{\rm{PTICM/P}}_{\rm{su}} = 5.5$ ns,
$t^{\rm{SA/Q}}_{\rm{su}} = t^{\rm{PTICM/Q}}_{\rm{su}} = 3.42$ ns,
$t^{\rm{SA/C}}_{\rm{su}} = t^{\rm{PTICM/C}}_{\rm{su}} = 2.6$ ns,
$t^{\rm{SQA_{\rm{ls}}/P}}_{su} =  1.08 \mu$s,
$t^{\rm{SQA_{\rm{ls}}/Q}}_{su} = 1.88 \mu$s,
$t^{\rm{SQA_{\rm{ls}}/C}}_{su} = 1.81 \mu$s,
$t^{\rm{SQA_{\rm{dws}}/C}}_{su} = 48.8 \mu$s. These times are used for
all figures with the exception of Fig.~\ref{f:onlyPTICM} and
Fig.~\ref{f:dw2x_qmc}. For the case of Fig.~\ref{f:onlyPTICM}, and to
give the best performance for each data set, we optimize for $R_{99}$
as described above for every instance of each of the CCFD and random
spin-glass data sets. Given that all the data sets are run with PTICM
and under the same computational resources, we plot directly the
wall-clock time required after the aforementioned optimization of
$MCS_{\rm{opt}}$.

To capture the computational scaling of SQA as a simulator of a
hypothetical quantum annealer [denoted as SQA(q)] we used the same
optimal values of $\rm{MCS}_{\rm{opt}}$ used for SQA but we do not
multiply by the factor of $N$. In this way we take into account the
intrinsic parallelism of quantum annealers. The prefactor $t_{\rm su}$
is changed as well to an arbitrary constant parameter, denoted
$t_{\rm{SQA(q)}}$ we can tune to make all the lines in
Fig.~\ref{f:dw2x_qmc} to have a similar $\rm TTS$ as that value obtained
by the DW2X device. The values used here were $t_{\rm{SQA(q)_{\rm{ls}}}}
= 5$ ns and $t_{\rm{SQA(q)_{\rm{dws}}}} = 1.3$ ns.

Because the SAT-based solver described below is significantly different
from the other stochastic solvers mentioned above, to estimate the $\rm
TTS$ we run the SAT-solver $1000$ times per instance and compute the
$\rm TTS$ for each run. From this distribution of $\rm TTS$ values, we
pick the 99\% percentile as the $\rm TTS$ value we report since it
matches the definition of the time needed to observe the desired
solution

\paragraph{SAT-based solver tailored for CCFD} --- The SAT-based
model-based diagnosis solver is implemented as follows. First it adds a
tree adder to the fault-augmented circuit to enforce the cardinality of
the fault. Second, the formula is converted to Conjunctive Normal Form
(CNF). Finally, a SAT solver is called $n$ times, first for computing
all zero-cardinality faults, then for all single faults, etc., until a
fault of cardinality $n$ is found. For our implementation we use
the highly-optimized SAT-solver \textsc{Lingeling}~\cite{biere16splatz}.
It is a deterministic SAT solver that uses Boolean
search enhancements, including symbolic optimization, occurrence lists,
literal stack, and clause distillation, etc.

\paragraph{DW2X programming details} --- When programming a quantum
annealer to solve real-world applications, the process of
minor embedding introduces many other parameters that do not exist when
benchmarking QA with a random spin-glass benchmark. One
common misconception is that implementing real-world applications is
harder because of the minor-embedding procedure. Although more efficient
embedding strategies are always desirable, we want to emphasize here
that it is not the main challenge when programming the device since
heuristic algorithms solve this problem reasonably well~\cite{Cai-14}.
It is also important to note here that the NP-hardness of finding the
smallest minor embedding (with respect to number of qubits) is largely
moot, because the smallest minor embedding is often far from optimal in
terms of performance. For example, from our experience, sometimes it is
preferable to have an embedding that uses more physical qubits but that
has shorter ``chains'' representing logical variables. In our work we
generate $100$ embeddings per instance regardless of the problem size.

The main challenge (the \emph{curse of limited
connectivity}~\cite{Benedetti2017,PerdomoOrtiz2017} due to quantum
annealers having a bounded number of couplers per qubit) does not lie in
the minor-embedding problem, but rather in the setting of the additional
parameters once the minor embedding has been chosen. Although proposals
exist to cope with this challenge
\cite{Choi2008,PerdomoOrtiz_arXiv2015a,Pudenz2016}, the optimal setting
of parameters is a largely open problem and one of the most important
ones affecting the performance of quantum annealers as optimizers
\cite{PerdomoOrtiz_arXiv2015a}. In this work, we use the strategy
proposed in Ref.~\cite{PerdomoOrtiz_arXiv2015a} to set the strength
$\Jferro$ of the ferromagnetic couplers, which enforce the
embedding, and for gauge selection.

In addition to setting $\Jferro$, we must also distribute the
logical biases $\{h_i\}$ and couplings $\{J_{i,j}\}$ over the available
physical biases and couplings $\{\tilde{h}_k\}$ and
$\{\tilde{J}_{k,l}\}$.  The key consideration in parameter setting is
the noise level of the programmable parameters of the quantum device.
The noise margin of the D-Wave 2X machine is $\tilde{h}_{j}<0.05$ for
biases and $\tilde{J}_{k,l}<0.1$ for couplers in a normalized,
hardware-embedded problem, with the difference due to the difference in
dynamic range.

We aim to divide the logical parameters as much as possible over the
corresponding physical parameters subject to this precision limit using
the following heuristic, which is similar to but distinct from that of
Ref.~\cite{Pudenz2016}. Consider a logical bias $h_i$ that corresponds
to $N_i$ hardware qubits. If $h_i / N_i$ is greater than the $0.05$
noise threshold, then each physical qubit $j$ is given a bias
$\tilde{h}_j = h_i / N_i$. If not, we consider the $n_i$ hardware qubits
within the chain that have nonzero interchain couplings.  If $h_i /n_i$
is greater than the threshold, we evenly distribute the logical bias
amongst these $n_i$ physical qubits. Finally, if neither of these
strategies exceed the threshold, we assign the logical bias completely
to hardware qubits with the lowest number of intrachain couplings,
breaking ties uniformly at random. The remaining hardware qubits within
the chain are given a bias of zero. We distribute the logical couplers
$J_{i,j}$ in a similar way. Suppose that the chains for logical qubits
$i$ and $j$ have $N_{i,j}$ physical couplers between them. If $J_{i,j} /
N_{i,j}$ is greater than $0.1$ (noise threshold), we evenly distribute
the logical coupling amongst the $N_{i,j}$ available physical couplers.
Otherwise, the logical coupling is completely assigned to a single
physical coupler uniformly at random from the $N_{i,j}$ options.

\section{Mapping of minimal-cardinality fault diagnosis for
combinational digital circuits to PUBO and QUBO}\label{s:mapping}

In this section we describe in detail two mappings of the fault-diagnosis problem to QUBO, via a mapping to PUBO. The original instance
consists of a set of $m$ gates, each with a specified hard fault model.
Excluding the inputs and outputs to the circuit, let $\mathbf x =
{(x_i)}_{i=1}^n \in {\{0, 1\}}^n$ indicate the value on every wire in
the circuit.  For gate $i$, let $\mathbf y_i \in {\{0, 1\}}^*$ be the
values of the input wires and $z_i$ the value of the output wire. These
are not new variables but rather alternative ways of referring to the
variables $\mathbf x$.  For example, if wire $i$ is the output of gate
$j$ and the first input into gate $k$, then $x_i$, $z_j$, and $y_{k, 1}$
all refer to the same variable. Let $g_i(\mathbf y_i) \in \{0, 1\}$ be
the Boolean function indicating the action of gate $i$, and $F_i(\mathbf
y_i, z_i) \in \{0, 1\}$ be the predicate indicating whether the combined
input $\mathbf y_i$ and output $z_i$ are consistent with the fault model
for gate $i$. Several examples for $g_i$ and $F_i$ are given in
Tables~\ref{tab:gate-polys} and~\ref{tab:fault-models}, respectively.

\begin{table}[!h]
\caption{Example gates and their representation as
polynomials. For details see the main text.}
\label{tab:gate-polys}
\begin{ruledtabular}
\begin{tabular}{lc} 
Gate & $g_i(\mathbf y_i)$  \\ \hline 
\OR{} & $y_{i,1} + y_{i, 2} - y_{i,1} y_{i,2}$ \\
\AND{} & $y_{i, 1} y_{i, 2}$ \\
\XOR{} & $y_{i,1} + y_{i, 2} - 2 y_{i,1} y_{i,2}$ \\
\EQ{} & $1 - y_{i, 1} - y_{i, 2} + 2 y_{i, 1} y_{i, 2}$ \\
\BUFFER{} & $y_{i,1}$ \\
\NOT{} & $1 - y_{i, 1}$ \\
\NOR{} & $1 - y_{i, 1} - y_{i, 2} + y_{i, 1} y_{i, 2}$ \\
\NAND{} & $1- y_{i,1} y_{i,2}$
\end{tabular}
\end{ruledtabular}
\end{table}

\begin{table}
\caption{Example fault models and their predicates as polynomials.
For details see the main text.}
\label{tab:fault-models}
\begin{ruledtabular}
\begin{tabular}{lc} 
Fault model & $F_i(\mathbf y_i, z_i)$ \\ \hline 
Stuck at $1$ & $z_i$ \\
Stuck at $0$ & $1-z_i$ \\
Stuck at $0$ or $1$ & $1$ \\
Stuck at first input & $\EQ(z_i, y_{i,1})$ \\
Stuck at first input or $0$ & $1 - y_{i,1} (1-z_i)$
\end{tabular}
\end{ruledtabular}
\end{table}

Bian {\em et al.}~\cite{Bian2016} have also used fault diagnosis as a
test bed for benchmarking novel techniques in QA.  They
used Satisfiability Modulo Theory to automatically generate functions
representing the cost function and constraints, whereas here we do so
manually, as described in this section. Their approach is further
differentiated from the present one by their use of problem
decomposition and locally structured embedding.

Note that we describe the mapping to pseudo-Boolean polynomials over
variables taking the values $\{0,1\}$, while the Hamiltonians in
physical quantum annealers directly represent functions of variables
taking the values $\pm 1$, i.e., Ising spins. The two representations
are equivalent with the following transformation:
\begin{align}
b &= (1 - s) / 2,
&
s &= 1 - 2b,
\end{align}
for $b \in \{0, 1\}$ and $s \in \{\pm 1\}$, with the latter being the
conventionally used for physical implementations on quantum annealers,
as in, e.g., Eq.~\eqref{eq:QUBO-C}. Note that the substitutions leave
the degree and connectivity of the polynomials unchanged.

\subsection{Explicit mapping}

For each gate $i$, introduce an additional variable $f_i$ that indicates
whether or not that gate is faulty. Assuming that $\mathbf f =
{(f_i)}_{i=1}^{N_{\rm{gates}}}$ is consistent with $\mathbf x_i$, the
number of faults is simply
\begin{equation}
\Hnumfault(\mathbf f) = \sum_{i=1}^{N_{\rm{gates}}} \Hnumfault^{(i)}(f_i)
=
\sum_{i=1}^{N_{\rm{gates}}} f_i.
\end{equation}
The consistency with the fault model is enforced by the penalty function
\begin{equation}
\begin{split}
\Hfaultset(\mathbf x, \mathbf f)
&=
\sum_{i=1}^{N_{\rm{gates}}}
\Hfaultset^{(i)}(\mathbf y_i, z_i, f_i),
\\
\Hfaultset^{(i)}(\mathbf y_i, z_i, f_i)
&=
\faultweight^{(i)}
    f_i \left[1 - F_i(\mathbf y_i, z_i)\right].
\end{split}
\end{equation}
Finally, we must also constrain the system to the appropriate behavior
when there is no fault:
\begin{equation}
\begin{split}
\Hgate(\mathbf x, \mathbf f)
&=
\sum_{i=1}^{N_{\rm{gates}}}
\Hgate^{(i)}(\mathbf x, \mathbf f), \\
\Hgate^{(i)}(\mathbf y_i, z_i, f_i)
&=
\gateweight^{(i)}
(1 - f_i) \XOR[g_i(\mathbf y_i), z_i].
\end{split}
\end{equation}
The overall cost function is 
\begin{equation}
\begin{split}
H(\mathbf x, \mathbf f)
&= \Hnumfault(\mathbf f) + \Hfaultset(\mathbf x, \mathbf f) + \Hgate(\mathbf x, \mathbf f)\\
&=
\sum_{i=1}^{N_{\rm{gates}}} H^{(i)}(\mathbf y_i, z_i, f_i),
\end{split}
\end{equation}
where
\begin{equation}
\begin{split}
H^{(i)}(\mathbf y_i, z_i, f_i)
&=
\Hnumfault(f_i) + 
\Hfaultset(\mathbf y_i, z_i, f_i) \\
&\hphantom{=\,}+
\Hgate(\mathbf y_i, z_i, f_i).
\end{split}
\end{equation}
Note that, in general, this function is quartic. Using two ancilla bits
per gate, the usual gadgets~\cite{perdomo08,Babbush2014} can be used to
reduce this to quadratic as needed. Depending on the circuit, some
ancilla bits may be reused to reduce the degree of the terms
corresponding to more than one gate. For example, if the input $\mathbf
y_i = (y_{i,1} y_{i,2})$ to gate $i$ happens to be the same input to
another gate $j$, then a single ancilla bit corresponding to $y_{i,1}
y_{i, 2}$ may be used for both gates. In this work, we use exactly two
ancilla bits per gate, corresponding to the conjunctions $y_{i,1}
y_{i,2}$ and $z_i f_i$.

The explicit mapping is easily extended to the case of $\nu >1$
input-output pairs.  Instead of the single $\mathbf x$, we have a copy
$\mathbf x_{\iota}$ for each input-output pair, and use a single set of
shared fault variables $\mathbf f$.  $\Hnumfault$ remains exactly the
same as above, while now there are copies of $\Hfaultset$ and $\Hgate$
for each input-output pair:
\begin{equation}
\begin{split}
\Hfaultset^{(i)}(\mathbf y_{i}, \mathbf z_i, f_i)
&= \sum_{\iota=1}^{\nu} 
\Hfaultset^{(i, \iota)}(\mathbf y_{i, \iota}, z_{i, \iota}, f_i), \\
\Hfaultset^{(i, \iota)}(\mathbf y_{i, \iota}, z_{i, \iota}, f_i)
&=
\faultweight^{(i)} 
f_i
    \left[1 - F_i(\mathbf y_{i, \iota}, z_{i, \iota})\right];
\end{split}
\end{equation}
and
\begin{equation}
\begin{split}
\Hgate^{(i)}(\mathbf y_i, \mathbf z_i, f_i)
&= \sum_{\iota=1}^{\nu}
    \Hgate^{(i, \iota)}(\mathbf y_{i, \iota}, z_{i, \iota}, f_i),\\
    \Hgate^{(i, \iota)}(\mathbf y_{i, \iota}, z_{i, \iota}, f_i)
&=
\gateweight^{(i)}
(1 - f_i) 
\XOR\left[g_i(\mathbf y_{i, \iota}), z_{i, \iota}\right];
\end{split}
\end{equation}
where $\mathbf y_{i,\iota}$ and $z_{i, \iota}$ are input and output bits
for gate $i$ in $\mathbf x_{\iota}$, and $\mathbf y_i$ and $\mathbf z_i$
contain all $\nu$ copies thereof.

The explicit mapping is also easily extended further to the case of $\mu
> 1$ fault modes.  For each gate $i$, we use $\mu$ fault variables
$\mathbf f_{i} = {\left(f_{i, \alpha}\right)}_{\alpha=1}^{\mu}$,
corresponding to the fault modes ${\left(F_{i, \alpha}
\right)}_{\alpha=1}^{\mu}$.  Considering $f_i = \sum_{\alpha=1}^{\mu}
f_{i, \alpha}$ as a function of $\mathbf f_i$ (rather than a separate
bit on its own), $\Hnumfault$ and $\Hgate$ remain unchanged from the
single-fault case, even with multiple input-output pairs.  Now there are
$\mu$ copies of $\Hfaultset$:
\begin{equation}
\begin{split}
\Hfaultset^{(i)}(\mathbf y_i, z_i, \mathbf f_i)
&=
\sum_{\iota=1}^{\nu} \sum_{\alpha = 1}^{\mu}
\Hfaultset^{(i,\iota, \alpha)}
    (\mathbf y_{i, \iota}, z_{i, \iota}, \mathbf f_{i, \alpha}),\\
\Hfaultset^{(i,\iota, \alpha)}
    (\mathbf y_{i, \iota}, z_{i, \iota}, \mathbf f_{i, \alpha})
&=
\faultweight^{(i)} 
f_{i, \alpha} 
    \left[1 - F_{i, \alpha} (\mathbf y_{i, \iota}, z_{i, \iota})\right].
\end{split}
\end{equation}
Finally, to penalize situations in which more than one fault bit is set
per gate, we add
\begin{equation}
\Hmultfault^{(i)} (\mathbf f_i)
= 
\multfaultweight^{(i)}
\sum_{\alpha=1}^{\mu-1} \sum_{\beta=\alpha + 1}^{\mu}
f_{i, \alpha} f_{i, \beta}.
\end{equation}
So long as $\multfaultweight^{(i)} > \nu \gateweight^{(i)}$,
$\Hmultfault$ outweighs the potentially negative $\Hgate$ as needed.
For each gate $i$, $\nu (1 + \mu)$ ancilla bits suffice, corresponding
to the conjunction of the bits $\mathbf y_{i, \iota}$ for each
input-output pair $\iota$ and to the conjunction $z_{i, \iota} f_{i,
\alpha}$ for every $\iota$ and mode $\alpha$.

When the fault modes considered are simply stuck at $1$ or stuck at $0$,
i.e. $F_i(\mathbf y_i, z_i) = F_i(z_i) = z_i$ or $1-z_i$, respectively,
we can use the alternative
\begin{equation}
\Hgate^{(i, \iota)}
=
\gateweight^{(i)}
\left\{1 + f_i [1 - 2F_i(z_{i, \iota})]\right\}
\XOR[g_i(\mathbf y_{i, \iota}), z_{i, \iota}],
\end{equation}
where $f_i = \sum_{\alpha=1}^{\mu}$ as before.  When $F_i$ is linear in
$z_i$, this expression is quadratic in $g_i$, $z_i$, and $f_i$, so that
it suffices to reduce $g_i$ to linear using a single ancilla bit
corresponding to the conjunction of the input bits $\mathbf y_i$.
Overall, only $\nu$ ancilla bits are needed per gate.

\subsection{Implicit mapping}\label{ss:implicit}

Having the fault bits $\mathbf f$ are not necessary.  Here we
show how to construct the requisite energy functions using just the wire
bits $\mathbf x$.  Note that $\Hfaultset$ is used only to enforce
consistency of the fault bits with the wire bits, and so is obviated by
the omission of the former.  Recall that we would like to find the
assignment of values to the wires that minimizes the number of faults
while being consistent with the nominal gates and fault models.
Therefore, we need a function $\Hnumfault^{(i)}$ that is zero when $z_i
= g_i(\mathbf y_i)$ and is one when $z_i \neq g_i(\mathbf y_i)$
\emph{and} $F_i(\mathbf y_i, z_i)$.  Its behavior when $z_i \neq
g_i(\mathbf y_i)$ and not $F_i(\mathbf y_i, z_i)$ only need be
non-negative; penalizing that case is left to $\Hgate$.  The following
meets our needs:
\begin{equation}
\begin{split}
\Hnumfault(\mathbf x) 
&= \sum_{i=1}^{N_{\rm{gates}}} \Hnumfault^{(i)}(\mathbf y_i, z_i) \\
&=
\sum_{i=1}^{N_{\rm{gates}}} F_i(\mathbf y_i, z_i) \XOR[g_i(\mathbf y_i), z_i].
\end{split}
\end{equation}
To penalize the case when the output $z_i$ of gate $i$ is inconsistent
with the input $\mathbf y_i$ but not in a way allowed by the fault
model, we use
\begin{equation}
\Hfaultset^{(i)}(\mathbf y_i, z_i)
=
\gateweight^{(i)}
[1 - F_i(\mathbf y_i, z_i)] \XOR[g_i(\mathbf y_i), z_i].
\end{equation}
The overall energy function for each gate is simply
\begin{equation}
H^{(i)}(\mathbf y_i, z_i)=
\Hnumfault^{(i)}(\mathbf y_i, z_i) + \Hgate^{(i)}(\mathbf y_i, z_i).
\end{equation}
Each $H^{(i)}$ is cubic, and can be reduced to quadratic using a single
ancilla bit.  As with the explicit mapping, in certain cases a single
ancilla may be shared among multiple gates.

For a single input-output pair, the implicit mapping naturally
generalizes to multiple fault modes, by considering a combined fault
mode that is the conjunction of the multiple ones, i.e., using $F_i =
\OR(F_{i, 1}, \ldots, F_{i, \mu})$.  Some examples, e.g., stuck at one
or first input, are shown in Table~\ref{tab:fault-models}.  This does
not apply to multiple input-output pairs because it does not enforce
that all copies are subject to the \emph{same} fault mode.  For
particular gates and sets of fault models, it is likely most efficient
to use a modification of the explicit mapping, as shown for the stuck at
$0$ and stuck at $1$ cases above.

\subsection{Logical penalty weights}

Without loss of generality, in this work we have chosen only one penalty
weight $\lambda$ for both $\gateweight^{(i)}$, which penalizes a
mismatch between the input and output of a gate in the absence of a
fault, and $\faultweight^{(i)}$, which enforces the fault model.  That
is, $\gateweight^{(i)} =\faultweight^{(i)}= \lambda $ for all $i$.
Setting $\lambda = N_{\rm{gates}}+1$ suffices to guarantee that the
global minima correspond to a valid diagnosis, i.e., those solutions
$(\mathbf x, \mathbf f)$ such that $\Hgate(\mathbf x, \mathbf f) =
\Hfaultset(\mathbf x, \mathbf f) = 0$.  Any valid diagnosis has energy
$H(\mathbf x, \mathbf f) = \Hnumfault$ at most $N_{\rm{gates}}$, so any
violation of the constraints incurring a penalty at least $\lambda =
N_{\mathrm{gates}} + 1$ yields a total energy greater than that of any
valid diagnosis.

A weaker condition to require of the penalty weight $\lambda$ is simply
that the ground state of $H$ is a valid diagnosis.  That is, an invalid
state (i.e., one that violates at least one of the model constraints)
may have lower total energy than \emph{some} valid state, but not than a
\emph{minimum}-fault valid state.  One simple upper bound on the minimum
number of faults is the number of outputs, which thus also serves as a
sufficient lower bound on $\lambda$.  In the case of the multiplier
circuits with $k$-bit and $l$-bit inputs, the length of the outputs in
simply $k+l$ bits, which is much smaller than $N_{\rm{gates}}$.

Nevertheless, a much lower value of $\lambda$ may suffice in practice
for a particular set of instances.  It is desirable to use the smallest
$\lambda$ possible, because when the coefficients of the Hamiltonian are
rescaled for a hardware implementation, larger values of $\lambda$ lead
to higher precision requirements, which may not be met by
limited-precision devices.  For the generation of the PUBO expressions
in the circuits considered here, up to mult8-8, we use a value of
$\lambda = 4$, regardless of the size of the circuit.  With the help of
the complete SAT-based solver, we check that, for all the instances
studied here, this value suffices to ensure that the ground state
corresponds to a valid diagnosis.

However, we did generate observations, not included in this study, for
which $\lambda = 4$ is insufficient.  This is extremely rare, from no
such examples in the smaller circuits to at most 1 in 500 for
the largest circuits.  Because we use the first hundred randomly
generated instances for each size, $\lambda = 4$ suffices for every
instance used; this is highly likely though not guaranteed.

A more common event that we had to filter in the instance generation was
the appearance of random instances where the minimal solution contains
no faults.  These are easy to eliminate since one can easily verify
whether the output corresponds to the multiplication of the inputs and
therefore the solution to our problem is trivial with a minimal fault
cardinality of zero.  It is interesting to note that in diagnosis task
such instances are still valuable, since one considers not only the
minimal cardinality but also the runners up could provide valuable
information about the circuit.  For example, it could be the case that
there is indeed a fault in the circuit but the output observations still
match the desired output, but the fault can only be unmasked for
example, by using another observation in the circuit.  The problem of
selecting the best inputs to probe faults in circuits is another
interesting NP-hard problems in its own.  We focus here on the minimal
cardinality case, given one input-output pairs.

\subsection{PUBO to QUBO reduction}

The cost function of the CCFD problem is initially expressed as a
pseudo-Boolean expression (i.e., PUBO) of degree greater than two.  We
then transform the higher-degree PUBO expression into a quadratic one by
using a conjunction gadget. The conjunction gadget introduces an ancilla
bit $q_{i, j}$ that corresponds to a conjunction of two bits $q_i$ and
$q_j$ in the PUBO, replaces all occurrences of the $q_i q_j$ with
$q_{i,j}$, and adds a penalty function so that in any ground state of
the QUBO expression the ancilla bit is appropriately set, $q_{i,j} = q_i
q_j$.  We use the penalty
function~\cite{perdomo08,babbush2013resource,Babbush2014}
\begin{equation}
H_{\mathrm{ancilla}}=
\delta (3q_{i,j}+ q_i q_j - 2q_i q_{i,j} -2q_j q_{i,j}),
\end{equation}
which is zero when $q_{i,j} = q_i q_j$ and at least $\delta$ otherwise,
where $\delta>0$ is the penalty weight.  The penalty weight $\delta$
needs to be large enough that states violating the ancilla constraint
have energy much larger than the ground energy of the original PUBO
expression, thus preserving the low-energy spectrum.  As with the
logical penalty $\lambda$, we would like $\delta$ to be as small as is
necessary in order to minimize the precision needed to implement the
cost function on a hardware device.  For each logic gate in the CCFD
problem, we determine that the following values are best, as a
multiple of the logical penalty weight $\lambda$:
\begin{equation}
\delta_{\AND}=\delta_{\OR}=2.5\lambda; \ \delta_{\XOR}=2\lambda.
\end{equation}

This controlled and optimized assignment of contraction penalties per logic gate is one of the remarkable features of this CCFD applications in contrast to others, where penalties can not only be higher but also scale with the number of variables~\cite{perdomo08,PerdomoOrtiz2012_LPF}. In this case, the penalties are independent of the circuit size.


\section{SQA vs DW2X}\label{s:dw2x_vs_qmc}

Here we address in more details the question of whether SQA has the same
scaling as the DW2X device and the comparison of the two schedules used
in the SQA simulations.  The linear schedule tends to underestimate the
scaling exponents for easy problems and small systems sizes, when the
required number of sweeps is small. This is because there might not be
enough QMC time to remove the segments in the imaginary time direction
that have different spin values.

For the DW schedule (dws) we cut the first 10\% of the schedule.  First,
the initial part of the D-Wave schedule is not necessary because it is
very easy to equilibrate QMC when the transverse-field strength is large
enough. Second, that leads to shorter simulation times as it takes
roughly the same time to run the first 10\% of the schedule as to run
the rest of the schedule.  This is because the SQA simulation time is
roughly proportional to the transverse-field strength and, in the first
part of the schedule, the transverse field is largest. Strictly
speaking, one probably can cut more than 10\% of the initial schedule.
One can also cut some fraction of the schedule at the end, but that will
not improve simulation times significantly.  However, that could lead
to a different scaling for easy problems and small small problem sizes.
This difference in scaling then could be fictitious and it might even
disappear for larger system sizes.

Therefore, it is difficult to make any conclusive statements about the
apparent difference in scaling and significant further work is required
to address this issue with more certainty.  Besides emphasizing that
such comparisons are not straightforward, these further simulations and
parameter fine tuning is beyond the scope of this work.

Note that the two statements are not contradictory with our statements about limited quantum speedup in Sec.~\ref{ss:physics-ccfd-scaling}. If we had unlimited computational resources we expect the SQA slopes to become smaller in value, while in the case of the DW2X we expect that optimization of the annealing time would lead to larger slope values compared to the current one. Although we declare the results of SQA vs DW2X inconclusive given the these two slopes might reach comparable values, given the expectation for SQA towards improving its scaling, these observations make our claims about limited quantum speedup even stronger.

\newpage

\newpage
\section{Qubit resources for numerical simulation and experiments}

\begin{figure}[!h]
\includegraphics[width=0.9\columnwidth]{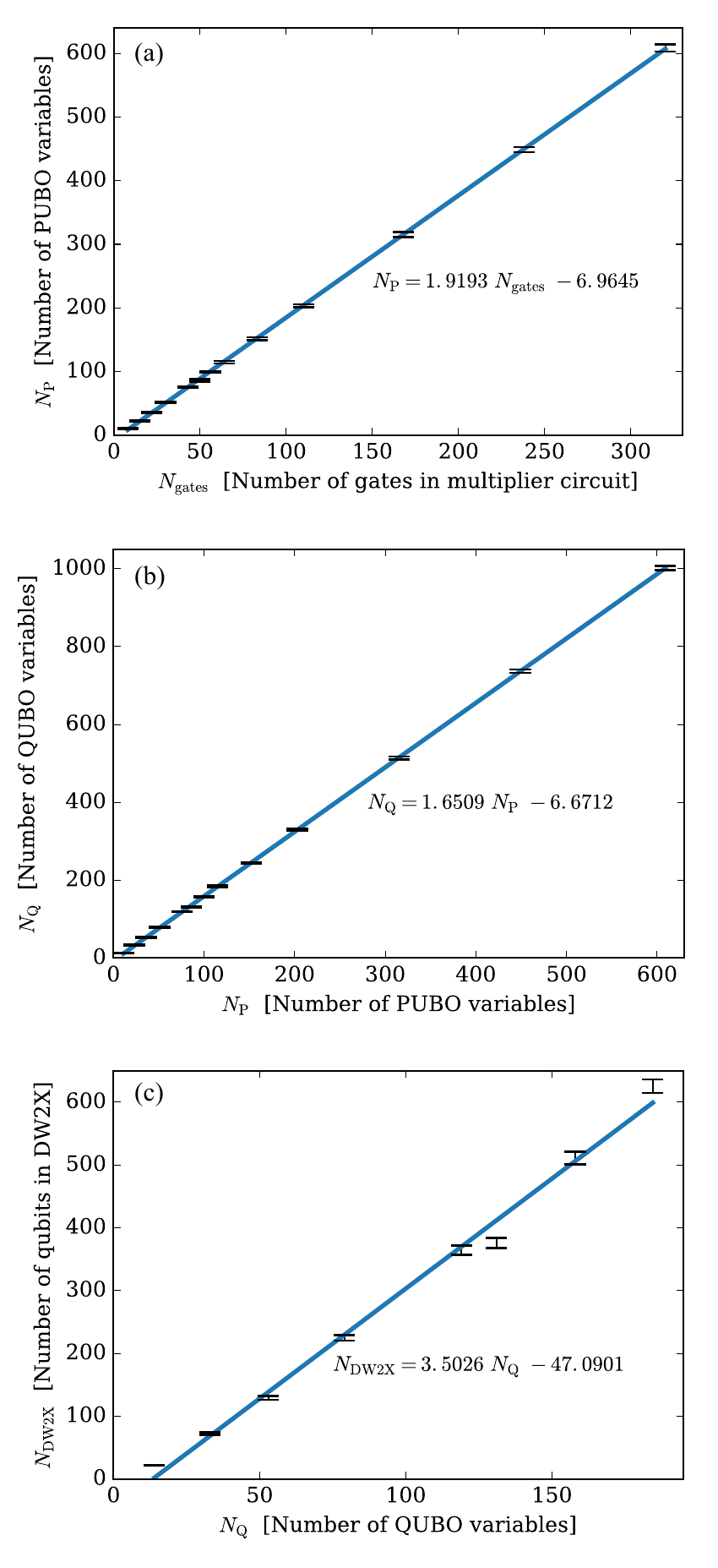}
\caption{Qubit resources for each of the problem representation [(a)
PUBO, (b) QUBO and (c) chimera (DW2X)] considered in our benchmarking
study of the CCFD instances. Data points correspond to the median values
extracted from a bootstrapping statistical analysis from $100$ instances
per problem size, with error bars indicating the 90\% confidence
intervals (CI).}
\label{f:qubit_resources}
\end{figure}

\onecolumngrid{}
\clearpage
\section{Intrinsic hardness of the CCFD instances compared to other
random spin-glass problems}\label{s:supp_hardness}

\begin{figure}[!h]
\includegraphics[width=0.95\textwidth]{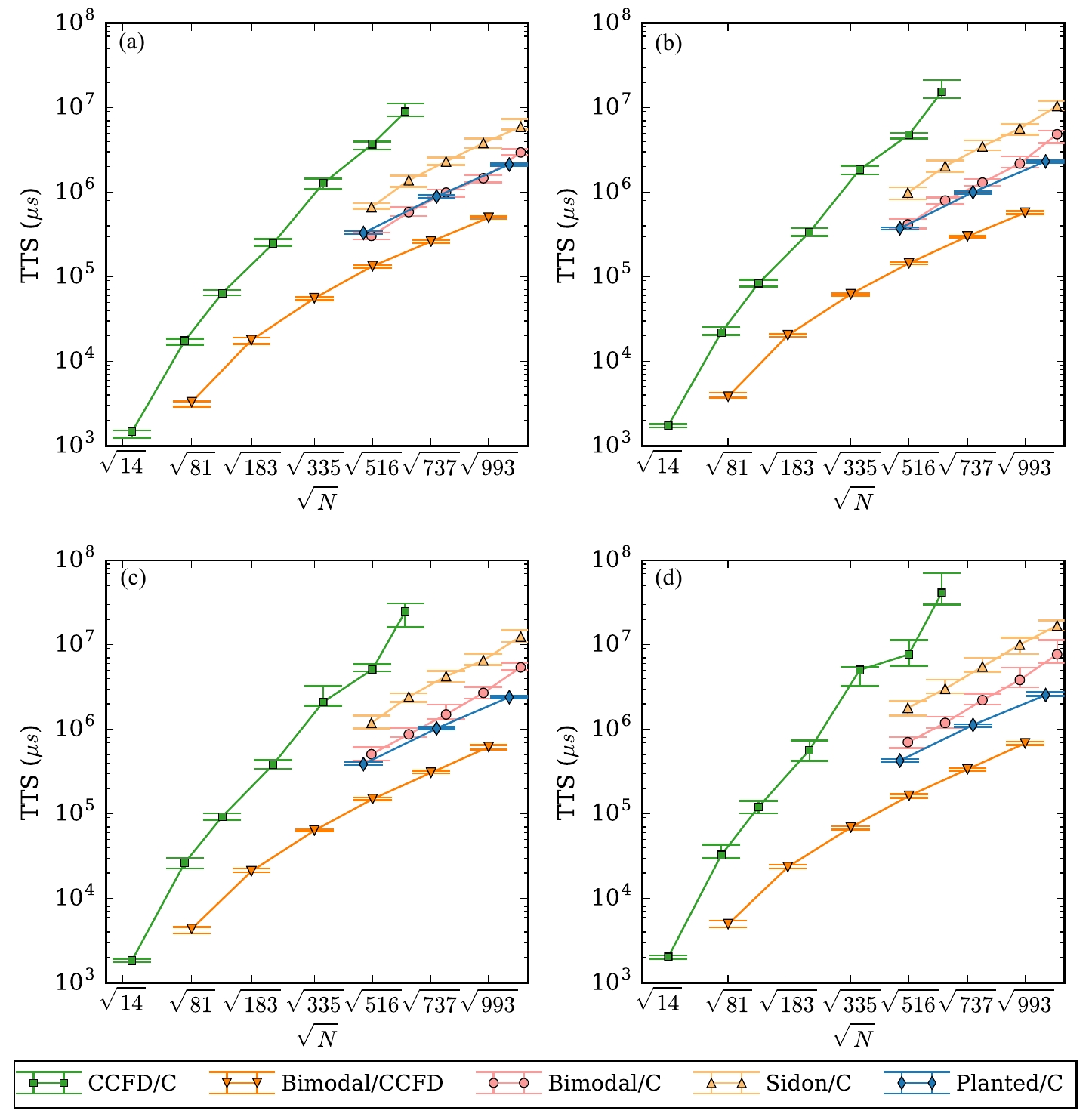}
\caption{Comparison of CCFD-based benchmark problems against other
random spin-glass benchmark classes, at different percentiles. Data
points correspond to the specific percentile value extracted from a
bootstrapping statistical analysis from $100$ instances per problem
size, with error bars indicating the 90\% CIs.}\label{f:supp_ICMonly_scaling}
\end{figure}

\newpage
\section{Scaling analysis from the application-centric perspective}

\begin{figure}[!h]
\includegraphics[width=0.95\textwidth]{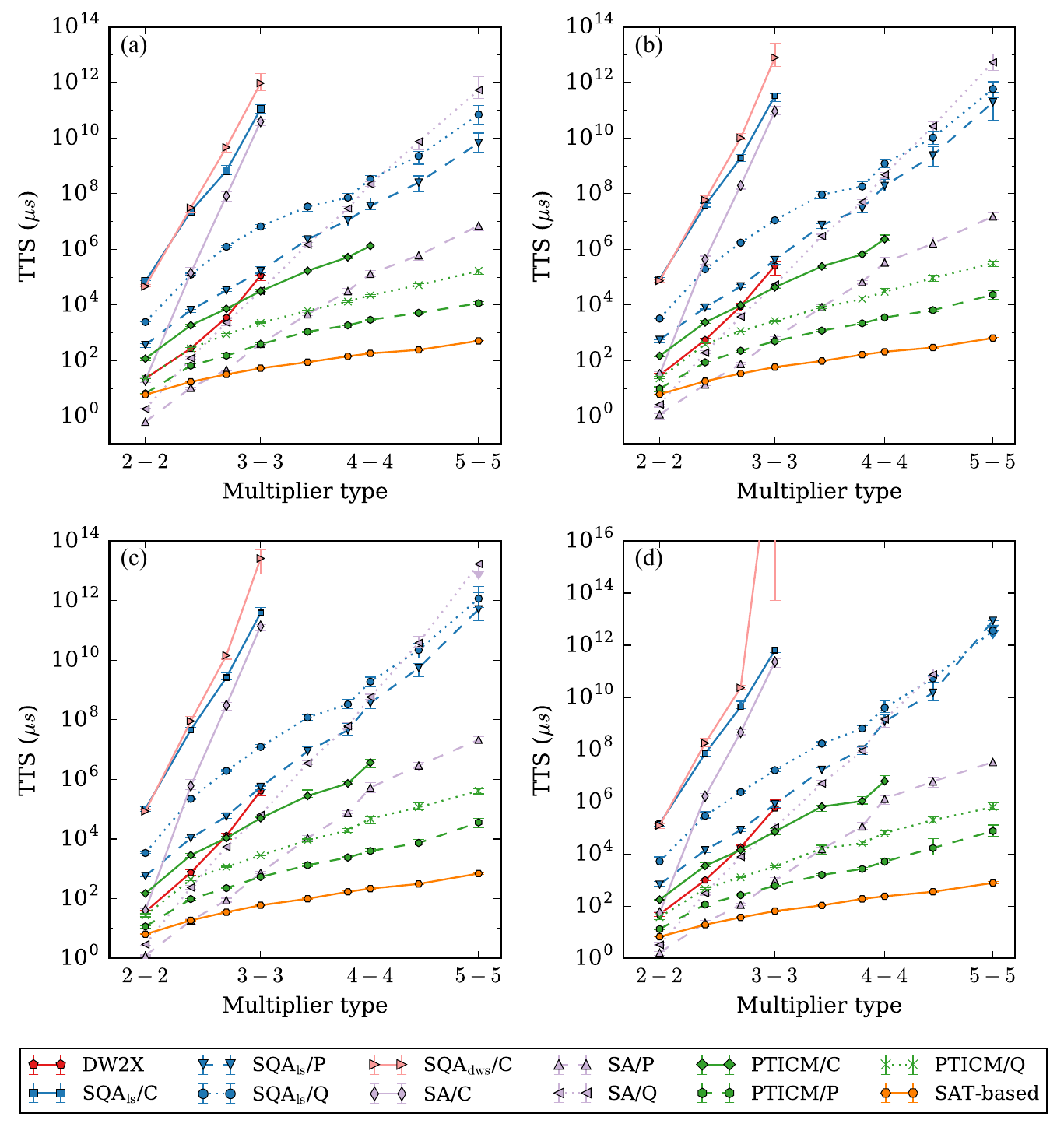}
\caption{Scaling analysis from the application-centric perspective at
different percentile levels.  (a) 25th, (b) 50th, (c) 60th, and (d) 75th
percentile.  Data points correspond to the specific percentile value
extracted from a bootstrapping statistical analysis from $100$ instances
per problem size, with error bars indicating the 90\% confidence
intervals (CI).}\label{f:supp_natural_scaling}

\end{figure}

\newpage
\section{Scaling analysis from the physics perspective}

\begin{figure}[!h]
\includegraphics[width=0.95\textwidth]{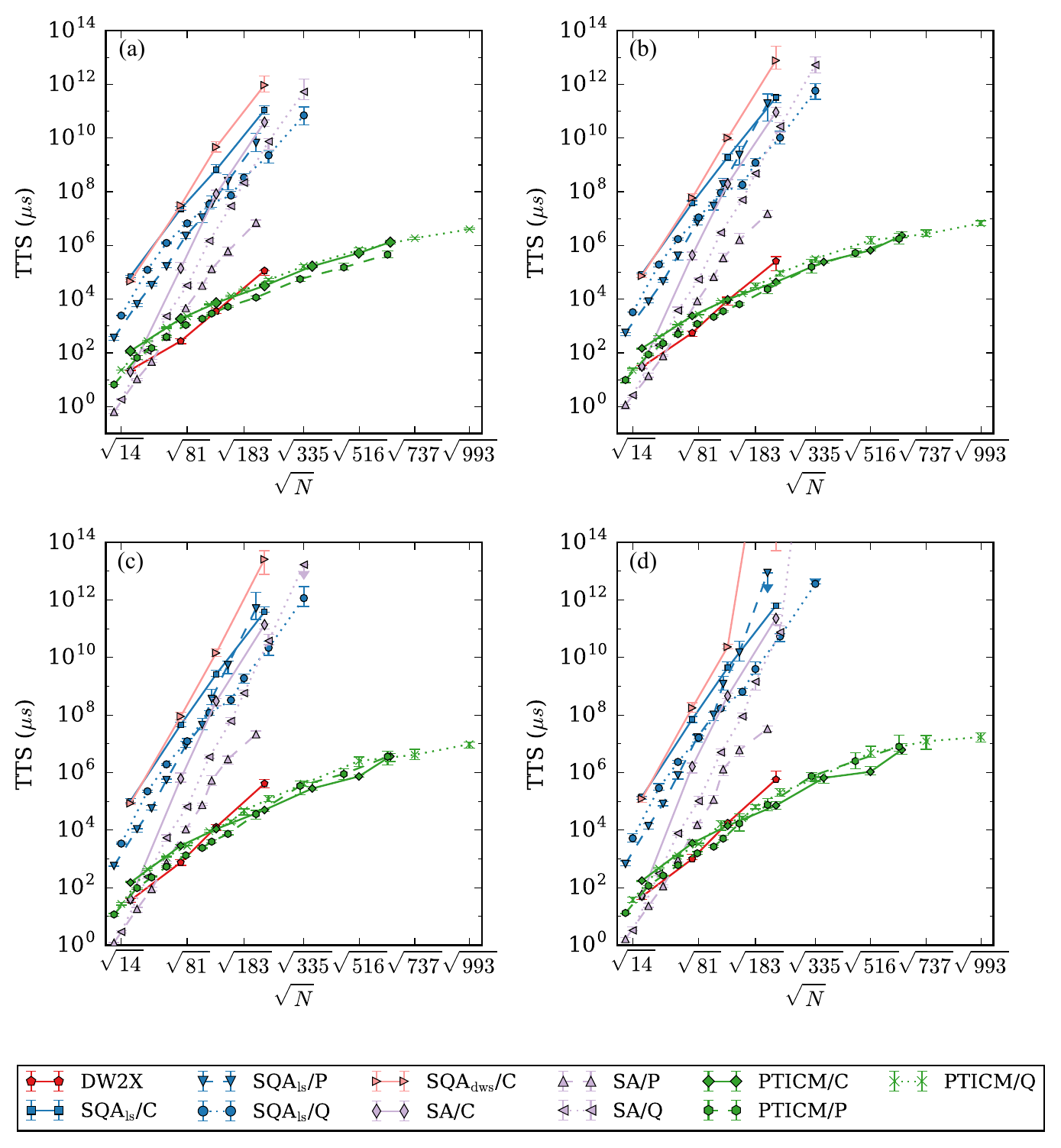}
\caption{Scaling analysis from the physics perspective at different
percentile levels. (a) 25th, (b) 50th, (c) 60th, and (d) 75th
percentile. Data points correspond to the specific percentile value
extracted from a bootstrapping statistical analysis from $100$ instances
per problem size, with error bars indicating the 90\% confidence
intervals (CI).}\label{f:supp_physics_scaling}
\end{figure}

\newpage
\twocolumngrid{}

\begin{thebibliography}{73}
\expandafter\ifx\csname natexlab\endcsname\relax\def\natexlab#1{#1}\fi
\expandafter\ifx\csname bibnamefont\endcsname\relax
  \def\bibnamefont#1{#1}\fi
\expandafter\ifx\csname bibfnamefont\endcsname\relax
  \def\bibfnamefont#1{#1}\fi
\expandafter\ifx\csname citenamefont\endcsname\relax
  \def\citenamefont#1{#1}\fi
\expandafter\ifx\csname url\endcsname\relax
  \def\url#1{\texttt{#1}}\fi
\expandafter\ifx\csname urlprefix\endcsname\relax\def\urlprefix{URL }\fi
\providecommand{\bibinfo}[2]{#2}
\providecommand{\eprint}[2][]{\url{#2}}

\bibitem[{\citenamefont{Kadowaki and Nishimori}(1998)}]{kadowaki:98}
\bibinfo{author}{\bibfnamefont{T.}~\bibnamefont{Kadowaki}} \bibnamefont{and}
  \bibinfo{author}{\bibfnamefont{H.}~\bibnamefont{Nishimori}},
  \emph{\bibinfo{title}{{{Quantum annealing in the transverse Ising model}}}},
  \bibinfo{journal}{Phys. Rev. E} \textbf{\bibinfo{volume}{58}},
  \bibinfo{pages}{5355} (\bibinfo{year}{1998}).

\bibitem[{\citenamefont{Finnila et~al.}(1994)\citenamefont{Finnila, Gomez,
  Sebenik, Stenson, and Doll}}]{finnila:94}
\bibinfo{author}{\bibfnamefont{A.~B.} \bibnamefont{Finnila}},
  \bibinfo{author}{\bibfnamefont{M.~A.} \bibnamefont{Gomez}},
  \bibinfo{author}{\bibfnamefont{C.}~\bibnamefont{Sebenik}},
  \bibinfo{author}{\bibfnamefont{C.}~\bibnamefont{Stenson}}, \bibnamefont{and}
  \bibinfo{author}{\bibfnamefont{J.~D.} \bibnamefont{Doll}},
  \emph{\bibinfo{title}{{{Quantum annealing: A new method for minimizing
  multidimensional functions}}}}, \bibinfo{journal}{Chem. Phys. Lett.}
  \textbf{\bibinfo{volume}{219}}, \bibinfo{pages}{343} (\bibinfo{year}{1994}).

\bibitem[{\citenamefont{Farhi et~al.}(2001)\citenamefont{Farhi, Goldstone,
  Gutmann, Lapan, Lundgren, and Preda}}]{farhi:01}
\bibinfo{author}{\bibfnamefont{E.}~\bibnamefont{Farhi}},
  \bibinfo{author}{\bibfnamefont{J.}~\bibnamefont{Goldstone}},
  \bibinfo{author}{\bibfnamefont{S.}~\bibnamefont{Gutmann}},
  \bibinfo{author}{\bibfnamefont{J.}~\bibnamefont{Lapan}},
  \bibinfo{author}{\bibfnamefont{A.}~\bibnamefont{Lundgren}}, \bibnamefont{and}
  \bibinfo{author}{\bibfnamefont{D.}~\bibnamefont{Preda}},
  \emph{\bibinfo{title}{{A} quantum adiabatic evolution algorithm applied to
  random instances of an {NP}-complete problem}}, \bibinfo{journal}{Science}
  \textbf{\bibinfo{volume}{292}}, \bibinfo{pages}{472} (\bibinfo{year}{2001}).

\bibitem[{\citenamefont{Santoro et~al.}(2002)\citenamefont{Santoro,
  Marto\v{n}\'ak, and Car}}]{santoro:02}
\bibinfo{author}{\bibfnamefont{G.}~\bibnamefont{Santoro}},
  \bibinfo{author}{\bibfnamefont{E.}~\bibnamefont{Marto\v{n}\'ak},
  \bibfnamefont{R.~Tosatti}}, \bibnamefont{and}
  \bibinfo{author}{\bibfnamefont{R.}~\bibnamefont{Car}},
  \emph{\bibinfo{title}{Theory of quantum annealing of an {I}sing spin glass}},
  \bibinfo{journal}{Science} \textbf{\bibinfo{volume}{295}},
  \bibinfo{pages}{2427} (\bibinfo{year}{2002}).

\bibitem[{\citenamefont{Das and Chakrabarti}(2005)}]{das:05}
\bibinfo{author}{\bibfnamefont{A.}~\bibnamefont{Das}} \bibnamefont{and}
  \bibinfo{author}{\bibfnamefont{B.~K.} \bibnamefont{Chakrabarti}},
  \emph{\bibinfo{title}{{{Quantum Annealing and Related Optimization
  Methods}}}} (\bibinfo{publisher}{Edited by A.~Das and B.K.~Chakrabarti,
  Lecture Notes in Physics 679, Berlin: Springer}, \bibinfo{year}{2005}).

\bibitem[{\citenamefont{Santoro and Tosatti}(2006)}]{santoro:06}
\bibinfo{author}{\bibfnamefont{G.~E.} \bibnamefont{Santoro}} \bibnamefont{and}
  \bibinfo{author}{\bibfnamefont{E.}~\bibnamefont{Tosatti}},
  \emph{\bibinfo{title}{{{TOPICAL REVIEW: Optimization using quantum mechanics:
  quantum annealing through adiabatic evolution}}}}, \bibinfo{journal}{J. Phys.
  A} \textbf{\bibinfo{volume}{39}}, \bibinfo{pages}{R393}
  (\bibinfo{year}{2006}).

\bibitem[{\citenamefont{Das and Chakrabarti}(2008)}]{das:08}
\bibinfo{author}{\bibfnamefont{A.}~\bibnamefont{Das}} \bibnamefont{and}
  \bibinfo{author}{\bibfnamefont{B.~K.} \bibnamefont{Chakrabarti}},
  \emph{\bibinfo{title}{{{Quantum Annealing and Analog Quantum Computation}}}},
  \bibinfo{journal}{Rev. Mod. Phys.} \textbf{\bibinfo{volume}{80}},
  \bibinfo{pages}{1061} (\bibinfo{year}{2008}).

\bibitem[{\citenamefont{Johnson et~al.}(2011)\citenamefont{Johnson, Amin,
  Gildert, Lanting, Hamze, Dickson, Harris, Berkley, Johansson, Bunyk
  et~al.}}]{johnson:11}
\bibinfo{author}{\bibfnamefont{M.~W.} \bibnamefont{Johnson}},
  \bibinfo{author}{\bibfnamefont{M.~H.~S.} \bibnamefont{Amin}},
  \bibinfo{author}{\bibfnamefont{S.}~\bibnamefont{Gildert}},
  \bibinfo{author}{\bibfnamefont{T.}~\bibnamefont{Lanting}},
  \bibinfo{author}{\bibfnamefont{F.}~\bibnamefont{Hamze}},
  \bibinfo{author}{\bibfnamefont{N.}~\bibnamefont{Dickson}},
  \bibinfo{author}{\bibfnamefont{R.}~\bibnamefont{Harris}},
  \bibinfo{author}{\bibfnamefont{A.~J.} \bibnamefont{Berkley}},
  \bibinfo{author}{\bibfnamefont{J.}~\bibnamefont{Johansson}},
  \bibinfo{author}{\bibfnamefont{P.}~\bibnamefont{Bunyk}},
  \bibnamefont{et~al.}, \emph{\bibinfo{title}{{Quantum annealing with
  manufactured spins}}}, \bibinfo{journal}{Nature}
  \textbf{\bibinfo{volume}{473}}, \bibinfo{pages}{194} (\bibinfo{year}{2011}).

\bibitem[{\citenamefont{{Dickson} et~al.}(2013)\citenamefont{{Dickson},
  {Johnson}, {Amin}, {Harris}, {Altomare}, {Berkley}, {Bunyk}, {Cai},
  {Chapple}, {Chavez} et~al.}}]{dickson:13}
\bibinfo{author}{\bibfnamefont{N.~G.} \bibnamefont{{Dickson}}},
  \bibinfo{author}{\bibfnamefont{M.~W.} \bibnamefont{{Johnson}}},
  \bibinfo{author}{\bibfnamefont{M.~H.} \bibnamefont{{Amin}}},
  \bibinfo{author}{\bibfnamefont{R.}~\bibnamefont{{Harris}}},
  \bibinfo{author}{\bibfnamefont{F.}~\bibnamefont{{Altomare}}},
  \bibinfo{author}{\bibfnamefont{A.~J.} \bibnamefont{{Berkley}}},
  \bibinfo{author}{\bibfnamefont{P.}~\bibnamefont{{Bunyk}}},
  \bibinfo{author}{\bibfnamefont{J.}~\bibnamefont{{Cai}}},
  \bibinfo{author}{\bibfnamefont{E.~M.} \bibnamefont{{Chapple}}},
  \bibinfo{author}{\bibfnamefont{P.}~\bibnamefont{{Chavez}}},
  \bibnamefont{et~al.}, \emph{\bibinfo{title}{{{Thermally assisted quantum
  annealing of a 16-qubit problem}}}}, \bibinfo{journal}{Nat. Commun.}
  \textbf{\bibinfo{volume}{4}}, \bibinfo{pages}{1903} (\bibinfo{year}{2013}).

\bibitem[{\citenamefont{{Boixo} et~al.}(2013)\citenamefont{{Boixo}, {Albash},
  {Spedalieri}, {Chancellor}, and {Lidar}}}]{boixo:13a}
\bibinfo{author}{\bibfnamefont{S.}~\bibnamefont{{Boixo}}},
  \bibinfo{author}{\bibfnamefont{T.}~\bibnamefont{{Albash}}},
  \bibinfo{author}{\bibfnamefont{F.~M.} \bibnamefont{{Spedalieri}}},
  \bibinfo{author}{\bibfnamefont{N.}~\bibnamefont{{Chancellor}}},
  \bibnamefont{and} \bibinfo{author}{\bibfnamefont{D.~A.}
  \bibnamefont{{Lidar}}}, \emph{\bibinfo{title}{{{Experimental signature of
  programmable quantum annealing}}}}, \bibinfo{journal}{Nat. Commun.}
  \textbf{\bibinfo{volume}{4}}, \bibinfo{pages}{2067} (\bibinfo{year}{2013}).

\bibitem[{\citenamefont{Katzgraber et~al.}(2014)\citenamefont{Katzgraber,
  Hamze, and Andrist}}]{katzgraber:14}
\bibinfo{author}{\bibfnamefont{H.~G.} \bibnamefont{Katzgraber}},
  \bibinfo{author}{\bibfnamefont{F.}~\bibnamefont{Hamze}}, \bibnamefont{and}
  \bibinfo{author}{\bibfnamefont{R.~S.} \bibnamefont{Andrist}},
  \emph{\bibinfo{title}{{Glassy Chimeras Could Be Blind to Quantum Speedup:
  Designing Better Benchmarks for Quantum Annealing Machines}}},
  \bibinfo{journal}{Phys. Rev. X} \textbf{\bibinfo{volume}{4}},
  \bibinfo{pages}{021008} (\bibinfo{year}{2014}).

\bibitem[{\citenamefont{{Boixo} et~al.}(2014)\citenamefont{{Boixo},
  {R{\o}nnow}, {Isakov}, {Wang}, {Wecker}, {Lidar}, {Martinis}, and
  {Troyer}}}]{boixo:14}
\bibinfo{author}{\bibfnamefont{S.}~\bibnamefont{{Boixo}}},
  \bibinfo{author}{\bibfnamefont{T.~F.} \bibnamefont{{R{\o}nnow}}},
  \bibinfo{author}{\bibfnamefont{S.~V.} \bibnamefont{{Isakov}}},
  \bibinfo{author}{\bibfnamefont{Z.}~\bibnamefont{{Wang}}},
  \bibinfo{author}{\bibfnamefont{D.}~\bibnamefont{{Wecker}}},
  \bibinfo{author}{\bibfnamefont{D.~A.} \bibnamefont{{Lidar}}},
  \bibinfo{author}{\bibfnamefont{J.~M.} \bibnamefont{{Martinis}}},
  \bibnamefont{and} \bibinfo{author}{\bibfnamefont{M.}~\bibnamefont{{Troyer}}},
  \emph{\bibinfo{title}{{Evidence for quantum annealing with more than one
  hundred qubits}}}, \bibinfo{journal}{Nat. Phys.}
  \textbf{\bibinfo{volume}{10}}, \bibinfo{pages}{218} (\bibinfo{year}{2014}).

\bibitem[{\citenamefont{{R{\o}nnow} et~al.}(2014)\citenamefont{{R{\o}nnow},
  {Wang}, {Job}, {Boixo}, {Isakov}, {Wecker}, {Martinis}, {Lidar}, and
  {Troyer}}}]{ronnow:14a}
\bibinfo{author}{\bibfnamefont{T.~F.} \bibnamefont{{R{\o}nnow}}},
  \bibinfo{author}{\bibfnamefont{Z.}~\bibnamefont{{Wang}}},
  \bibinfo{author}{\bibfnamefont{J.}~\bibnamefont{{Job}}},
  \bibinfo{author}{\bibfnamefont{S.}~\bibnamefont{{Boixo}}},
  \bibinfo{author}{\bibfnamefont{S.~V.} \bibnamefont{{Isakov}}},
  \bibinfo{author}{\bibfnamefont{D.}~\bibnamefont{{Wecker}}},
  \bibinfo{author}{\bibfnamefont{J.~M.} \bibnamefont{{Martinis}}},
  \bibinfo{author}{\bibfnamefont{D.~A.} \bibnamefont{{Lidar}}},
  \bibnamefont{and} \bibinfo{author}{\bibfnamefont{M.}~\bibnamefont{{Troyer}}},
  \emph{\bibinfo{title}{{Defining and detecting quantum speedup}}},
  \bibinfo{journal}{Science} \textbf{\bibinfo{volume}{345}},
  \bibinfo{pages}{420} (\bibinfo{year}{2014}).

\bibitem[{\citenamefont{Katzgraber et~al.}(2015)\citenamefont{Katzgraber,
  Hamze, Zhu, Ochoa, and Munoz-Bauza}}]{katzgraber:15}
\bibinfo{author}{\bibfnamefont{H.~G.} \bibnamefont{Katzgraber}},
  \bibinfo{author}{\bibfnamefont{F.}~\bibnamefont{Hamze}},
  \bibinfo{author}{\bibfnamefont{Z.}~\bibnamefont{Zhu}},
  \bibinfo{author}{\bibfnamefont{A.~J.} \bibnamefont{Ochoa}}, \bibnamefont{and}
  \bibinfo{author}{\bibfnamefont{H.}~\bibnamefont{Munoz-Bauza}},
  \emph{\bibinfo{title}{{Seeking Quantum Speedup Through Spin Glasses: The
  Good, the Bad, and the Ugly}}}, \bibinfo{journal}{Phys. Rev. X}
  \textbf{\bibinfo{volume}{5}}, \bibinfo{pages}{031026} (\bibinfo{year}{2015}).

\bibitem[{\citenamefont{Heim et~al.}(2015)\citenamefont{Heim, R{\o}nnow,
  Isakov, and Troyer}}]{heim:15}
\bibinfo{author}{\bibfnamefont{B.}~\bibnamefont{Heim}},
  \bibinfo{author}{\bibfnamefont{T.~F.} \bibnamefont{R{\o}nnow}},
  \bibinfo{author}{\bibfnamefont{S.~V.} \bibnamefont{Isakov}},
  \bibnamefont{and} \bibinfo{author}{\bibfnamefont{M.}~\bibnamefont{Troyer}},
  \emph{\bibinfo{title}{{Quantum versus classical annealing of Ising spin
  glasses}}}, \bibinfo{journal}{Science} \textbf{\bibinfo{volume}{348}},
  \bibinfo{pages}{215} (\bibinfo{year}{2015}).

\bibitem[{\citenamefont{Hen et~al.}(2015)\citenamefont{Hen, Job, Albash,
  R{\o}nnow, Troyer, and Lidar}}]{hen:15a}
\bibinfo{author}{\bibfnamefont{I.}~\bibnamefont{Hen}},
  \bibinfo{author}{\bibfnamefont{J.}~\bibnamefont{Job}},
  \bibinfo{author}{\bibfnamefont{T.}~\bibnamefont{Albash}},
  \bibinfo{author}{\bibfnamefont{T.~F.} \bibnamefont{R{\o}nnow}},
  \bibinfo{author}{\bibfnamefont{M.}~\bibnamefont{Troyer}}, \bibnamefont{and}
  \bibinfo{author}{\bibfnamefont{D.~A.} \bibnamefont{Lidar}},
  \emph{\bibinfo{title}{{Probing for quantum speedup in spin-glass problems
  with planted solutions}}}, \bibinfo{journal}{Phys. Rev. A}
  \textbf{\bibinfo{volume}{92}}, \bibinfo{pages}{042325}
  (\bibinfo{year}{2015}).

\bibitem[{\citenamefont{Rieffel et~al.}(2015)\citenamefont{Rieffel, Venturelli,
  O'Gorman, Do, Prystay, and Smelyanskiy}}]{rieffel:15}
\bibinfo{author}{\bibfnamefont{E.~G.} \bibnamefont{Rieffel}},
  \bibinfo{author}{\bibfnamefont{D.}~\bibnamefont{Venturelli}},
  \bibinfo{author}{\bibfnamefont{B.}~\bibnamefont{O'Gorman}},
  \bibinfo{author}{\bibfnamefont{M.~B.} \bibnamefont{Do}},
  \bibinfo{author}{\bibfnamefont{E.~M.} \bibnamefont{Prystay}},
  \bibnamefont{and} \bibinfo{author}{\bibfnamefont{V.~N.}
  \bibnamefont{Smelyanskiy}}, \emph{\bibinfo{title}{A case study in programming
  a quantum annealer for hard operational planning problems}},
  \bibinfo{journal}{Quant. Inf. Proc.} \textbf{\bibinfo{volume}{14}},
  \bibinfo{pages}{1} (\bibinfo{year}{2015}).

\bibitem[{\citenamefont{{Boixo} et~al.}(2016)\citenamefont{{Boixo},
  {Smelyanskiy}, {Shabani}, {Isakov}, {Dykman}, {Denchev}, {Amin}, {Smirnov},
  {Mohseni}, and {Neven}}}]{boixo:16}
\bibinfo{author}{\bibfnamefont{S.}~\bibnamefont{{Boixo}}},
  \bibinfo{author}{\bibfnamefont{V.~N.} \bibnamefont{{Smelyanskiy}}},
  \bibinfo{author}{\bibfnamefont{A.}~\bibnamefont{{Shabani}}},
  \bibinfo{author}{\bibfnamefont{S.~V.} \bibnamefont{{Isakov}}},
  \bibinfo{author}{\bibfnamefont{M.}~\bibnamefont{{Dykman}}},
  \bibinfo{author}{\bibfnamefont{V.~S.} \bibnamefont{{Denchev}}},
  \bibinfo{author}{\bibfnamefont{M.~H.} \bibnamefont{{Amin}}},
  \bibinfo{author}{\bibfnamefont{A.~Y.} \bibnamefont{{Smirnov}}},
  \bibinfo{author}{\bibfnamefont{M.}~\bibnamefont{{Mohseni}}},
  \bibnamefont{and} \bibinfo{author}{\bibfnamefont{H.}~\bibnamefont{{Neven}}},
  \emph{\bibinfo{title}{{Computational multiqubit tunnelling in programmable
  quantum annealers}}}, \bibinfo{journal}{Nat. Comm.}
  \textbf{\bibinfo{volume}{7}}, \bibinfo{pages}{10327} (\bibinfo{year}{2016}).

\bibitem[{\citenamefont{Denchev et~al.}(2016)\citenamefont{Denchev, Boixo,
  Isakov, Ding, Babbush, Smelyanskiy, Martinis, and Neven}}]{denchev:16}
\bibinfo{author}{\bibfnamefont{V.~S.} \bibnamefont{Denchev}},
  \bibinfo{author}{\bibfnamefont{S.}~\bibnamefont{Boixo}},
  \bibinfo{author}{\bibfnamefont{S.~V.} \bibnamefont{Isakov}},
  \bibinfo{author}{\bibfnamefont{N.}~\bibnamefont{Ding}},
  \bibinfo{author}{\bibfnamefont{R.}~\bibnamefont{Babbush}},
  \bibinfo{author}{\bibfnamefont{V.}~\bibnamefont{Smelyanskiy}},
  \bibinfo{author}{\bibfnamefont{J.}~\bibnamefont{Martinis}}, \bibnamefont{and}
  \bibinfo{author}{\bibfnamefont{H.}~\bibnamefont{Neven}},
  \emph{\bibinfo{title}{{W}hat is the {C}omputational {V}alue of {F}inite
  {R}ange {T}unneling?}}, \bibinfo{journal}{Phys. Rev. X}
  \textbf{\bibinfo{volume}{6}}, \bibinfo{pages}{031015} (\bibinfo{year}{2016}).

\bibitem[{\citenamefont{King et~al.}(2019)\citenamefont{King, Yarkoni, Raymond,
  Ozfidan, King, Nevisi, Hilton, and McGeoch}}]{king:19}
\bibinfo{author}{\bibfnamefont{J.}~\bibnamefont{King}},
  \bibinfo{author}{\bibfnamefont{S.}~\bibnamefont{Yarkoni}},
  \bibinfo{author}{\bibfnamefont{J.}~\bibnamefont{Raymond}},
  \bibinfo{author}{\bibfnamefont{I.}~\bibnamefont{Ozfidan}},
  \bibinfo{author}{\bibfnamefont{A.~D.} \bibnamefont{King}},
  \bibinfo{author}{\bibfnamefont{M.~M.} \bibnamefont{Nevisi}},
  \bibinfo{author}{\bibfnamefont{J.~P.} \bibnamefont{Hilton}},
  \bibnamefont{and} \bibinfo{author}{\bibfnamefont{C.~C.}
  \bibnamefont{McGeoch}}, \emph{\bibinfo{title}{Quantum annealing amid local
  ruggedness and global frustration}}, \bibinfo{journal}{Journal of the
  Physical Society of Japan} \textbf{\bibinfo{volume}{88}},
  \bibinfo{pages}{061007} (\bibinfo{year}{2019}).

\bibitem[{\citenamefont{{Mandr{\`a}} et~al.}(2016)\citenamefont{{Mandr{\`a}},
  {Zhu}, {Wang}, {Perdomo-Ortiz}, and {Katzgraber}}}]{mandra:16b}
\bibinfo{author}{\bibfnamefont{S.}~\bibnamefont{{Mandr{\`a}}}},
  \bibinfo{author}{\bibfnamefont{Z.}~\bibnamefont{{Zhu}}},
  \bibinfo{author}{\bibfnamefont{W.}~\bibnamefont{{Wang}}},
  \bibinfo{author}{\bibfnamefont{A.}~\bibnamefont{{Perdomo-Ortiz}}},
  \bibnamefont{and} \bibinfo{author}{\bibfnamefont{H.~G.}
  \bibnamefont{{Katzgraber}}}, \emph{\bibinfo{title}{{Strengths and weaknesses
  of weak-strong cluster problems: A detailed overview of state-of-the-art
  classical heuristics versus quantum approaches}}}, \bibinfo{journal}{Phys.
  Rev. A} \textbf{\bibinfo{volume}{94}}, \bibinfo{pages}{022337}
  (\bibinfo{year}{2016}).

\bibitem[{\citenamefont{Perdomo et~al.}(2008)\citenamefont{Perdomo, Truncik,
  {Tubert-Brohman}, Rose, and {Aspuru-Guzik}}}]{perdomo08}
\bibinfo{author}{\bibfnamefont{A.}~\bibnamefont{Perdomo}},
  \bibinfo{author}{\bibfnamefont{C.}~\bibnamefont{Truncik}},
  \bibinfo{author}{\bibfnamefont{I.}~\bibnamefont{{Tubert-Brohman}}},
  \bibinfo{author}{\bibfnamefont{G.}~\bibnamefont{Rose}}, \bibnamefont{and}
  \bibinfo{author}{\bibfnamefont{A.}~\bibnamefont{{Aspuru-Guzik}}},
  \emph{\bibinfo{title}{Construction of model hamiltonians for adiabatic
  quantum computation and its application to finding low-energy conformations
  of lattice protein models}}, \bibinfo{journal}{Phys. Rev. A}
  \textbf{\bibinfo{volume}{78}}, \bibinfo{pages}{012320}
  (\bibinfo{year}{2008}).

\bibitem[{\citenamefont{Perdomo-Ortiz et~al.}(2012)\citenamefont{Perdomo-Ortiz,
  Dickson, Drew-Brook, Rose, and Aspuru-Guzik}}]{PerdomoOrtiz2012_LPF}
\bibinfo{author}{\bibfnamefont{A.}~\bibnamefont{Perdomo-Ortiz}},
  \bibinfo{author}{\bibfnamefont{N.}~\bibnamefont{Dickson}},
  \bibinfo{author}{\bibfnamefont{M.}~\bibnamefont{Drew-Brook}},
  \bibinfo{author}{\bibfnamefont{G.}~\bibnamefont{Rose}}, \bibnamefont{and}
  \bibinfo{author}{\bibfnamefont{A.}~\bibnamefont{Aspuru-Guzik}},
  \emph{\bibinfo{title}{Finding low-energy conformations of lattice protein
  models by quantum annealing}}, \bibinfo{journal}{Sci. Rep.}
  \textbf{\bibinfo{volume}{2}}, \bibinfo{pages}{571} (\bibinfo{year}{2012}).

\bibitem[{\citenamefont{Gaitan and Clark}(2012)}]{Gaitan2012}
\bibinfo{author}{\bibfnamefont{F.}~\bibnamefont{Gaitan}} \bibnamefont{and}
  \bibinfo{author}{\bibfnamefont{L.}~\bibnamefont{Clark}},
  \emph{\bibinfo{title}{Ramsey numbers and adiabatic quantum computing}},
  \bibinfo{journal}{Phys. Rev. Lett.} \textbf{\bibinfo{volume}{108}},
  \bibinfo{pages}{010501} (\bibinfo{year}{2012}).

\bibitem[{\citenamefont{{O'Gorman, B.} et~al.}(2015)\citenamefont{{O'Gorman,
  B.}, {Babbush, R.}, {Perdomo-Ortiz, A.}, {Aspuru-Guzik, A.}, and
  {Smelyanskiy, V.}}}]{OGormanEPJST2015}
\bibinfo{author}{\bibnamefont{{O'Gorman, B.}}},
  \bibinfo{author}{\bibnamefont{{Babbush, R.}}},
  \bibinfo{author}{\bibnamefont{{Perdomo-Ortiz, A.}}},
  \bibinfo{author}{\bibnamefont{{Aspuru-Guzik, A.}}}, \bibnamefont{and}
  \bibinfo{author}{\bibnamefont{{Smelyanskiy, V.}}},
  \emph{\bibinfo{title}{Bayesian network structure learning using quantum
  annealing}}, \bibinfo{journal}{Eur. Phys. J. Special Topics}
  \textbf{\bibinfo{volume}{224}}, \bibinfo{pages}{163} (\bibinfo{year}{2015}).

\bibitem[{\citenamefont{Perdomo-Ortiz
  et~al.}(2015{\natexlab{a}})\citenamefont{Perdomo-Ortiz, Fluegemann,
  Narasimhan, Biswas, and Smelyanskiy}}]{PerdomoOrtiz_EPJST2015}
\bibinfo{author}{\bibfnamefont{A.}~\bibnamefont{Perdomo-Ortiz}},
  \bibinfo{author}{\bibfnamefont{J.}~\bibnamefont{Fluegemann}},
  \bibinfo{author}{\bibfnamefont{S.}~\bibnamefont{Narasimhan}},
  \bibinfo{author}{\bibfnamefont{R.}~\bibnamefont{Biswas}}, \bibnamefont{and}
  \bibinfo{author}{\bibfnamefont{V.~N.} \bibnamefont{Smelyanskiy}},
  \emph{\bibinfo{title}{A quantum annealing approach for fault detection and
  diagnosis of graph-based systems}}, \bibinfo{journal}{Eur. Phys. J. Special
  Topics} \textbf{\bibinfo{volume}{224}}, \bibinfo{pages}{131}
  (\bibinfo{year}{2015}{\natexlab{a}}).

\bibitem[{\citenamefont{Zick et~al.}(2015)\citenamefont{Zick, Shehab, and
  French}}]{Zick2015}
\bibinfo{author}{\bibfnamefont{K.~M.} \bibnamefont{Zick}},
  \bibinfo{author}{\bibfnamefont{O.}~\bibnamefont{Shehab}}, \bibnamefont{and}
  \bibinfo{author}{\bibfnamefont{M.}~\bibnamefont{French}},
  \emph{\bibinfo{title}{Experimental quantum annealing: case study involving
  the graph isomorphism problem}}, \bibinfo{journal}{Scientific Reports}
  \textbf{\bibinfo{volume}{5}}, \bibinfo{pages}{11168 EP }
  (\bibinfo{year}{2015}).

\bibitem[{\citenamefont{Neukart et~al.}(2017)\citenamefont{Neukart,
  Compostella, Seidel, von Dollen, Yarkoni, and Parney}}]{Neukart2017}
\bibinfo{author}{\bibfnamefont{F.}~\bibnamefont{Neukart}},
  \bibinfo{author}{\bibfnamefont{G.}~\bibnamefont{Compostella}},
  \bibinfo{author}{\bibfnamefont{C.}~\bibnamefont{Seidel}},
  \bibinfo{author}{\bibfnamefont{D.}~\bibnamefont{von Dollen}},
  \bibinfo{author}{\bibfnamefont{S.}~\bibnamefont{Yarkoni}}, \bibnamefont{and}
  \bibinfo{author}{\bibfnamefont{B.}~\bibnamefont{Parney}},
  \emph{\bibinfo{title}{Traffic flow optimization using a quantum annealer}},
  \bibinfo{journal}{arXiv:1708.01625v2}  (\bibinfo{year}{2017}).

\bibitem[{\citenamefont{Bian et~al.}(2016)\citenamefont{Bian, Chudak, Israel,
  Lackey, Macready, and Roy}}]{Bian2016}
\bibinfo{author}{\bibfnamefont{Z.}~\bibnamefont{Bian}},
  \bibinfo{author}{\bibfnamefont{F.}~\bibnamefont{Chudak}},
  \bibinfo{author}{\bibfnamefont{R.~B.} \bibnamefont{Israel}},
  \bibinfo{author}{\bibfnamefont{B.}~\bibnamefont{Lackey}},
  \bibinfo{author}{\bibfnamefont{W.~G.} \bibnamefont{Macready}},
  \bibnamefont{and} \bibinfo{author}{\bibfnamefont{A.}~\bibnamefont{Roy}},
  \emph{\bibinfo{title}{Mapping constrained optimization problems to quantum
  annealing with application to fault diagnosis}}, \bibinfo{journal}{Frontiers
  in ICT} \textbf{\bibinfo{volume}{3}}, \bibinfo{pages}{14}
  (\bibinfo{year}{2016}).

\bibitem[{\citenamefont{Feldman et~al.}(2010)\citenamefont{Feldman, Provan, and
  van Gemund}}]{feldman10safari}
\bibinfo{author}{\bibfnamefont{A.}~\bibnamefont{Feldman}},
  \bibinfo{author}{\bibfnamefont{G.}~\bibnamefont{Provan}}, \bibnamefont{and}
  \bibinfo{author}{\bibfnamefont{A.}~\bibnamefont{van Gemund}},
  \emph{\bibinfo{title}{Approximate model-based diagnosis using greedy
  stochastic search}}, \bibinfo{journal}{Journal of Artificial Intelligence
  Research} \textbf{\bibinfo{volume}{38}}, \bibinfo{pages}{371}
  (\bibinfo{year}{2010}).

\bibitem[{\citenamefont{Feldman et~al.}(2007)\citenamefont{Feldman, Provan, and
  van Gemund}}]{Feldman2007}
\bibinfo{author}{\bibfnamefont{A.}~\bibnamefont{Feldman}},
  \bibinfo{author}{\bibfnamefont{G.}~\bibnamefont{Provan}}, \bibnamefont{and}
  \bibinfo{author}{\bibfnamefont{A.}~\bibnamefont{van Gemund}}, in
  \emph{\bibinfo{booktitle}{Abstraction, Reformulation, and Approximation: 7th
  International Symposium, SARA 2007, Whistler, Canada, July 18-21, 2007.
  Proceedings}}, edited by
  \bibinfo{editor}{\bibfnamefont{I.}~\bibnamefont{Miguel}} \bibnamefont{and}
  \bibinfo{editor}{\bibfnamefont{W.}~\bibnamefont{Ruml}}
  (\bibinfo{publisher}{Springer Berlin Heidelberg}, \bibinfo{address}{Berlin,
  Heidelberg}, \bibinfo{year}{2007}), pp. \bibinfo{pages}{139--154}.

\bibitem[{\citenamefont{Eiter and Gottlob}(1995)}]{eiter95complexity}
\bibinfo{author}{\bibfnamefont{T.}~\bibnamefont{Eiter}} \bibnamefont{and}
  \bibinfo{author}{\bibfnamefont{G.}~\bibnamefont{Gottlob}},
  \emph{\bibinfo{title}{The complexity of logic-based abduction}},
  \bibinfo{journal}{Journal of the ACM} \textbf{\bibinfo{volume}{42}},
  \bibinfo{pages}{3} (\bibinfo{year}{1995}).

\bibitem[{\citenamefont{{Rieger, H.} and {Kawashima,
  N.}}(1999)}]{rieger:kawashima}
\bibinfo{author}{\bibnamefont{{Rieger, H.}}} \bibnamefont{and}
  \bibinfo{author}{\bibnamefont{{Kawashima, N.}}},
  \emph{\bibinfo{title}{Application of a continuous time cluster algorithm to
  the two-dimensional random quantum ising ferromagnet}},
  \bibinfo{journal}{Eur. Phys. J. B} \textbf{\bibinfo{volume}{9}},
  \bibinfo{pages}{233} (\bibinfo{year}{1999}).

\bibitem[{\citenamefont{Kirkpatrick et~al.}(1983)\citenamefont{Kirkpatrick,
  {Gelatt, Jr.}, and Vecchi}}]{kirkpatrick:83}
\bibinfo{author}{\bibfnamefont{S.}~\bibnamefont{Kirkpatrick}},
  \bibinfo{author}{\bibfnamefont{C.~D.} \bibnamefont{{Gelatt, Jr.}}},
  \bibnamefont{and} \bibinfo{author}{\bibfnamefont{M.~P.}
  \bibnamefont{Vecchi}}, \emph{\bibinfo{title}{Optimization by simulated
  annealing}}, \bibinfo{journal}{Science} \textbf{\bibinfo{volume}{220}},
  \bibinfo{pages}{671} (\bibinfo{year}{1983}).

\bibitem[{\citenamefont{{Isakov} et~al.}(2015)\citenamefont{{Isakov},
  {Zintchenko}, {R{\o}nnow}, and {Troyer}}}]{isakov:15}
\bibinfo{author}{\bibfnamefont{S.~V.} \bibnamefont{{Isakov}}},
  \bibinfo{author}{\bibfnamefont{I.~N.} \bibnamefont{{Zintchenko}}},
  \bibinfo{author}{\bibfnamefont{T.~F.} \bibnamefont{{R{\o}nnow}}},
  \bibnamefont{and} \bibinfo{author}{\bibfnamefont{M.}~\bibnamefont{{Troyer}}},
  \emph{\bibinfo{title}{{{Optimized simulated annealing for Ising spin
  glasses}}}}, \bibinfo{journal}{Comput. Phys. Commun.}
  \textbf{\bibinfo{volume}{192}}, \bibinfo{pages}{265} (\bibinfo{year}{2015}),
  \bibinfo{note}{(see also ancillary material to arxiv:cond-mat/1401.1084)}.

\bibitem[{\citenamefont{Hukushima and Nemoto}(1996)}]{hukushima:96}
\bibinfo{author}{\bibfnamefont{K.}~\bibnamefont{Hukushima}} \bibnamefont{and}
  \bibinfo{author}{\bibfnamefont{K.}~\bibnamefont{Nemoto}},
  \emph{\bibinfo{title}{Exchange {M}onte {C}arlo method and application to spin
  glass simulations}}, \bibinfo{journal}{J. Phys. Soc. Jpn.}
  \textbf{\bibinfo{volume}{65}}, \bibinfo{pages}{1604} (\bibinfo{year}{1996}).

\bibitem[{\citenamefont{Katzgraber et~al.}(2006)\citenamefont{Katzgraber,
  Trebst, Huse, and Troyer}}]{katzgraber:06a}
\bibinfo{author}{\bibfnamefont{H.~G.} \bibnamefont{Katzgraber}},
  \bibinfo{author}{\bibfnamefont{S.}~\bibnamefont{Trebst}},
  \bibinfo{author}{\bibfnamefont{D.~A.} \bibnamefont{Huse}}, \bibnamefont{and}
  \bibinfo{author}{\bibfnamefont{M.}~\bibnamefont{Troyer}},
  \emph{\bibinfo{title}{{{Feedback-optimized parallel tempering Monte
  Carlo}}}}, \bibinfo{journal}{J. Stat. Mech.}
  \textbf{\bibinfo{volume}{\normalfont{P03018}}} (\bibinfo{year}{2006}).

\bibitem[{\citenamefont{Moreno et~al.}(2003)\citenamefont{Moreno, Katzgraber,
  and Hartmann}}]{moreno:03}
\bibinfo{author}{\bibfnamefont{J.~J.} \bibnamefont{Moreno}},
  \bibinfo{author}{\bibfnamefont{H.~G.} \bibnamefont{Katzgraber}},
  \bibnamefont{and} \bibinfo{author}{\bibfnamefont{A.~K.}
  \bibnamefont{Hartmann}}, \emph{\bibinfo{title}{Finding low-temperature states
  with parallel tempering, simulated annealing and simple {M}onte {C}arlo}},
  \bibinfo{journal}{Int. J. Mod. Phys. C} \textbf{\bibinfo{volume}{14}},
  \bibinfo{pages}{285} (\bibinfo{year}{2003}).

\bibitem[{\citenamefont{{Zhu} et~al.}(2015)\citenamefont{{Zhu}, {Ochoa}, and
  {Katzgraber}}}]{zhu:15b}
\bibinfo{author}{\bibfnamefont{Z.}~\bibnamefont{{Zhu}}},
  \bibinfo{author}{\bibfnamefont{A.~J.} \bibnamefont{{Ochoa}}},
  \bibnamefont{and} \bibinfo{author}{\bibfnamefont{H.~G.}
  \bibnamefont{{Katzgraber}}}, \emph{\bibinfo{title}{{{Efficient Cluster
  Algorithm for Spin Glasses in Any Space Dimension}}}},
  \bibinfo{journal}{Phys. Rev. Lett.} \textbf{\bibinfo{volume}{115}},
  \bibinfo{pages}{077201} (\bibinfo{year}{2015}).

\bibitem[{\citenamefont{Mandr\`a et~al.}(2016)\citenamefont{Mandr\`a, Zhu,
  Wang, Perdomo-Ortiz, and Katzgraber}}]{mandra:16}
\bibinfo{author}{\bibfnamefont{S.}~\bibnamefont{Mandr\`a}},
  \bibinfo{author}{\bibfnamefont{Z.}~\bibnamefont{Zhu}},
  \bibinfo{author}{\bibfnamefont{W.}~\bibnamefont{Wang}},
  \bibinfo{author}{\bibfnamefont{A.}~\bibnamefont{Perdomo-Ortiz}},
  \bibnamefont{and} \bibinfo{author}{\bibfnamefont{H.~G.}
  \bibnamefont{Katzgraber}}, \emph{\bibinfo{title}{Strengths and weaknesses of
  weak-strong cluster problems: A detailed overview of state-of-the-art
  classical heuristics versus quantum approaches}}, \bibinfo{journal}{Phys.
  Rev. A} \textbf{\bibinfo{volume}{94}}, \bibinfo{pages}{022337}
  (\bibinfo{year}{2016}).

\bibitem[{\citenamefont{Lucas}(2014)}]{lucas:14}
\bibinfo{author}{\bibfnamefont{A.}~\bibnamefont{Lucas}},
  \emph{\bibinfo{title}{{Ising formulations of many NP problems}}},
  \bibinfo{journal}{Front. Physics} \textbf{\bibinfo{volume}{12}},
  \bibinfo{pages}{5} (\bibinfo{year}{2014}).

\bibitem[{\citenamefont{Chancellor et~al.}(2017)\citenamefont{Chancellor,
  Zohren, and Warburton}}]{Chancellor2017}
\bibinfo{author}{\bibfnamefont{N.}~\bibnamefont{Chancellor}},
  \bibinfo{author}{\bibfnamefont{S.}~\bibnamefont{Zohren}}, \bibnamefont{and}
  \bibinfo{author}{\bibfnamefont{P.~A.} \bibnamefont{Warburton}},
  \emph{\bibinfo{title}{Circuit design for multi-body interactions in
  superconducting quantum annealing systems with applications to a scalable
  architecture}}, \bibinfo{journal}{npj Quantum Information}
  \textbf{\bibinfo{volume}{3}}, \bibinfo{pages}{21} (\bibinfo{year}{2017}).

\bibitem[{\citenamefont{Boros and Hammer}(2002)}]{Boros2002}
\bibinfo{author}{\bibfnamefont{E.}~\bibnamefont{Boros}} \bibnamefont{and}
  \bibinfo{author}{\bibfnamefont{P.~L.} \bibnamefont{Hammer}},
  \emph{\bibinfo{title}{Pseudo-boolean optimization}},
  \bibinfo{journal}{Discrete Appl. Math.} \textbf{\bibinfo{volume}{123}},
  \bibinfo{pages}{155} (\bibinfo{year}{2002}).

\bibitem[{\citenamefont{Choi}(2011)}]{Choi2011}
\bibinfo{author}{\bibfnamefont{V.}~\bibnamefont{Choi}},
  \emph{\bibinfo{title}{Minor-embedding in adiabatic quantum computation: {II}.
  minor-universal graph design}}, \bibinfo{journal}{Quantum Information
  Processing} \textbf{\bibinfo{volume}{10}}, \bibinfo{pages}{343}
  (\bibinfo{year}{2011}), ISSN \bibinfo{issn}{1570-0755}.

\bibitem[{\citenamefont{Cai et~al.}(2014)\citenamefont{Cai, Macready, and
  Roy}}]{Cai-14}
\bibinfo{author}{\bibfnamefont{J.}~\bibnamefont{Cai}},
  \bibinfo{author}{\bibfnamefont{B.}~\bibnamefont{Macready}}, \bibnamefont{and}
  \bibinfo{author}{\bibfnamefont{A.}~\bibnamefont{Roy}},
  \emph{\bibinfo{title}{A practical heuristic for finding graph minors}},
  \bibinfo{journal}{arXiv:1406.2741}  (\bibinfo{year}{2014}).

\bibitem[{\citenamefont{Perdomo-Ortiz
  et~al.}(2015{\natexlab{b}})\citenamefont{Perdomo-Ortiz, Fluegemann, Biswas,
  and Smelyanskiy}}]{PerdomoOrtiz_arXiv2015a}
\bibinfo{author}{\bibfnamefont{A.}~\bibnamefont{Perdomo-Ortiz}},
  \bibinfo{author}{\bibfnamefont{J.}~\bibnamefont{Fluegemann}},
  \bibinfo{author}{\bibfnamefont{R.}~\bibnamefont{Biswas}}, \bibnamefont{and}
  \bibinfo{author}{\bibfnamefont{V.~N.} \bibnamefont{Smelyanskiy}},
  \emph{\bibinfo{title}{A performance estimator for quantum annealers: Gauge
  selection and parameter setting}}, \bibinfo{journal}{arXiv:1503.01083}
  (\bibinfo{year}{2015}{\natexlab{b}}).

\bibitem[{\citenamefont{Zhu et~al.}(2016)\citenamefont{Zhu, Ochoa, Hamze,
  Schnabel, and Katzgraber}}]{zhu:16}
\bibinfo{author}{\bibfnamefont{Z.}~\bibnamefont{Zhu}},
  \bibinfo{author}{\bibfnamefont{A.~J.} \bibnamefont{Ochoa}},
  \bibinfo{author}{\bibfnamefont{F.}~\bibnamefont{Hamze}},
  \bibinfo{author}{\bibfnamefont{S.}~\bibnamefont{Schnabel}}, \bibnamefont{and}
  \bibinfo{author}{\bibfnamefont{H.~G.} \bibnamefont{Katzgraber}},
  \emph{\bibinfo{title}{{{Best-case performance of quantum annealers on native
  spin-glass benchmarks: How chaos can affect success probabilities}}}},
  \bibinfo{journal}{Phys. Rev. A} \textbf{\bibinfo{volume}{93}},
  \bibinfo{pages}{012317} (\bibinfo{year}{2016}).

\bibitem[{\citenamefont{Sidon}(1932)}]{sidon:32}
\bibinfo{author}{\bibfnamefont{S.}~\bibnamefont{Sidon}},
  \emph{\bibinfo{title}{{{Ein Satz {\"u}ber trigonometrische Polynome und seine
  Anwendung in der Theorie der Fourier-Reihen}}}},
  \bibinfo{journal}{Mathematische Annalen} \textbf{\bibinfo{volume}{106}},
  \bibinfo{pages}{536} (\bibinfo{year}{1932}).

\bibitem[{\citenamefont{Selby}(2014)}]{Selby2014}
\bibinfo{author}{\bibfnamefont{A.}~\bibnamefont{Selby}},
  \emph{\bibinfo{title}{Efficient subgraph-based sampling of isingtype models
  with frustration}}, \bibinfo{journal}{arXiv:1409.3934}
  (\bibinfo{year}{2014}).

\bibitem[{\citenamefont{Albash and Lidar}(2017)}]{Albash2017}
\bibinfo{author}{\bibfnamefont{T.}~\bibnamefont{Albash}} \bibnamefont{and}
  \bibinfo{author}{\bibfnamefont{D.~A.} \bibnamefont{Lidar}},
  \emph{\bibinfo{title}{Evidence for a limited quantum speedup on a quantum
  annealer}}, \bibinfo{journal}{arXiv:1705.07452}  (\bibinfo{year}{2017}).

\bibitem[{\citenamefont{Job and Lidar}(2018)}]{Job2018}
\bibinfo{author}{\bibfnamefont{J.}~\bibnamefont{Job}} \bibnamefont{and}
  \bibinfo{author}{\bibfnamefont{D.}~\bibnamefont{Lidar}},
  \emph{\bibinfo{title}{Test-driving 1000 qubits}}, \bibinfo{journal}{Quantum
  Science and Technology} \textbf{\bibinfo{volume}{3}}, \bibinfo{pages}{030501}
  (\bibinfo{year}{2018}).

\bibitem[{\citenamefont{Aaronson}(2015{\natexlab{a}})}]{AaronsonBlogGooglePaper}
\bibinfo{author}{\bibfnamefont{S.}~\bibnamefont{Aaronson}},
  \emph{\bibinfo{title}{Google, d-wave, and the case of the factor - 10**8
  speedup for what?}},
  \bibinfo{journal}{http://www.scottaaronson.com/blog/?p=2555}
  (\bibinfo{year}{2015}{\natexlab{a}}).

\bibitem[{\citenamefont{Aaronson}(2015{\natexlab{b}})}]{AaronsonBlogDWbenchmarks}
\bibinfo{author}{\bibfnamefont{S.}~\bibnamefont{Aaronson}},
  \emph{\bibinfo{title}{Insert d-wave post here}},
  \bibinfo{journal}{http://www.scottaaronson.com/blog/?p=3192}
  (\bibinfo{year}{2015}{\natexlab{b}}).

\bibitem[{\citenamefont{Hormozi et~al.}(2017)\citenamefont{Hormozi, Brown,
  Carleo, and Troyer}}]{Hormozi2016}
\bibinfo{author}{\bibfnamefont{L.}~\bibnamefont{Hormozi}},
  \bibinfo{author}{\bibfnamefont{E.~W.} \bibnamefont{Brown}},
  \bibinfo{author}{\bibfnamefont{G.}~\bibnamefont{Carleo}}, \bibnamefont{and}
  \bibinfo{author}{\bibfnamefont{M.}~\bibnamefont{Troyer}},
  \emph{\bibinfo{title}{Nonstoquastic hamiltonians and quantum annealing of an
  ising spin glass}}, \bibinfo{journal}{Phys. Rev. B}
  \textbf{\bibinfo{volume}{95}}, \bibinfo{pages}{184416}
  (\bibinfo{year}{2017}).

\bibitem[{\citenamefont{Nishimori and Takada}(2017)}]{Nishimori2017}
\bibinfo{author}{\bibfnamefont{H.}~\bibnamefont{Nishimori}} \bibnamefont{and}
  \bibinfo{author}{\bibfnamefont{K.}~\bibnamefont{Takada}},
  \emph{\bibinfo{title}{Exponential enhancement of the efficiency of quantum
  annealing by non-stoquastic hamiltonians}}, \bibinfo{journal}{Frontiers in
  ICT} \textbf{\bibinfo{volume}{4}}, \bibinfo{pages}{2} (\bibinfo{year}{2017}).

\bibitem[{\citenamefont{Biere}(2016)}]{biere16splatz}
\bibinfo{author}{\bibfnamefont{A.}~\bibnamefont{Biere}},
  \emph{\bibinfo{title}{\textsc{Splatz}, \textsc{Lingeling},
  \textsc{Plingeling}, \textsc{Treengeling}, \textsc{YalSAT} entering the {SAT}
  competition 2016}}, \bibinfo{journal}{SAT COMPETITION 2016}
  p.~\bibinfo{pages}{44} (\bibinfo{year}{2016}).

\bibitem[{\citenamefont{Strand et~al.}(2017)\citenamefont{Strand, Przybysz,
  Ferguson, and Zick}}]{Strand2017}
\bibinfo{author}{\bibfnamefont{J.}~\bibnamefont{Strand}},
  \bibinfo{author}{\bibfnamefont{A.}~\bibnamefont{Przybysz}},
  \bibinfo{author}{\bibfnamefont{D.}~\bibnamefont{Ferguson}}, \bibnamefont{and}
  \bibinfo{author}{\bibfnamefont{K.}~\bibnamefont{Zick}},
  \emph{\bibinfo{title}{Zzz coupler for native embedding of max-3sat problem
  instances in quantum annealing hardware}}, \bibinfo{journal}{Bulletin of the
  American Physical Society}  (\bibinfo{year}{2017}).

\bibitem[{\citenamefont{{Perdomo-Ortiz}
  et~al.}(2010)\citenamefont{{Perdomo-Ortiz}, {Venegas-Andraca}, and
  {Aspuru-Guzik}}}]{PerdomoSAQC2010}
\bibinfo{author}{\bibfnamefont{A.}~\bibnamefont{{Perdomo-Ortiz}}},
  \bibinfo{author}{\bibfnamefont{S.~E.} \bibnamefont{{Venegas-Andraca}}},
  \bibnamefont{and}
  \bibinfo{author}{\bibfnamefont{A.}~\bibnamefont{{Aspuru-Guzik}}},
  \emph{\bibinfo{title}{A study of heuristic guesses for adiabatic quantum
  computation}}, \bibinfo{journal}{Quantum Inf. Process.}
  \textbf{\bibinfo{volume}{10}}, \bibinfo{pages}{33} (\bibinfo{year}{2010}),
  ISSN \bibinfo{issn}{1570-0755, 1573-1332}.

\bibitem[{\citenamefont{Chancellor}(2017{\natexlab{a}})}]{Chancellor2017a}
\bibinfo{author}{\bibfnamefont{N.}~\bibnamefont{Chancellor}},
  \emph{\bibinfo{title}{Modernizing quantum annealing using local searches}},
  \bibinfo{journal}{New Journal of Physics} \textbf{\bibinfo{volume}{19}},
  \bibinfo{pages}{023024} (\bibinfo{year}{2017}{\natexlab{a}}).

\bibitem[{\citenamefont{Chancellor}(2017{\natexlab{b}})}]{Chancellor2017b}
\bibinfo{author}{\bibfnamefont{N.}~\bibnamefont{Chancellor}},
  \emph{\bibinfo{title}{Modernizing quantum annealing ii: Genetic algorithms
  and inference}}, \bibinfo{journal}{arXiv:1609.05875}
  (\bibinfo{year}{2017}{\natexlab{b}}).

\bibitem[{\citenamefont{Karimi et~al.}(2017)\citenamefont{Karimi, Rosenberg,
  and Katzgraber}}]{Karimi2017}
\bibinfo{author}{\bibfnamefont{H.}~\bibnamefont{Karimi}},
  \bibinfo{author}{\bibfnamefont{G.}~\bibnamefont{Rosenberg}},
  \bibnamefont{and} \bibinfo{author}{\bibfnamefont{H.~G.}
  \bibnamefont{Katzgraber}}, \emph{\bibinfo{title}{Effective optimization using
  sample persistence: A case study on quantum annealers and various monte carlo
  optimization methods}}, \bibinfo{journal}{Phys. Rev. E}
  \textbf{\bibinfo{volume}{96}}, \bibinfo{pages}{043312}
  (\bibinfo{year}{2017}).

\bibitem[{\citenamefont{Harris et~al.}(2010)\citenamefont{Harris, Johnson,
  Lanting, Berkley, Johansson, Bunyk, Tolkacheva, Ladizinsky, Ladizinsky, Oh
  et~al.}}]{harris2010}
\bibinfo{author}{\bibfnamefont{R.}~\bibnamefont{Harris}},
  \bibinfo{author}{\bibfnamefont{M.~W.} \bibnamefont{Johnson}},
  \bibinfo{author}{\bibfnamefont{T.}~\bibnamefont{Lanting}},
  \bibinfo{author}{\bibfnamefont{A.~J.} \bibnamefont{Berkley}},
  \bibinfo{author}{\bibfnamefont{J.}~\bibnamefont{Johansson}},
  \bibinfo{author}{\bibfnamefont{P.}~\bibnamefont{Bunyk}},
  \bibinfo{author}{\bibfnamefont{E.}~\bibnamefont{Tolkacheva}},
  \bibinfo{author}{\bibfnamefont{E.}~\bibnamefont{Ladizinsky}},
  \bibinfo{author}{\bibfnamefont{N.}~\bibnamefont{Ladizinsky}},
  \bibinfo{author}{\bibfnamefont{T.}~\bibnamefont{Oh}}, \bibnamefont{et~al.},
  \emph{\bibinfo{title}{Experimental investigation of an eight-qubit unit cell
  in a superconducting optimization processor}}, \bibinfo{journal}{Phys. Rev.
  B.} \textbf{\bibinfo{volume}{82}}, \bibinfo{pages}{024511}
  (\bibinfo{year}{2010}).

\bibitem[{\citenamefont{Barahona}(1982)}]{Barahona1982}
\bibinfo{author}{\bibfnamefont{F.}~\bibnamefont{Barahona}},
  \emph{\bibinfo{title}{On the computational complexity of ising spin glass
  models}}, \bibinfo{journal}{Journal of Physics A: Mathematical and General}
  \textbf{\bibinfo{volume}{15}}, \bibinfo{pages}{3241} (\bibinfo{year}{1982}).

\bibitem[{\citenamefont{Albash et~al.}(2012)\citenamefont{Albash, Boixo, Lidar,
  and Zanardi}}]{AlbashNJP2012}
\bibinfo{author}{\bibfnamefont{T.}~\bibnamefont{Albash}},
  \bibinfo{author}{\bibfnamefont{S.}~\bibnamefont{Boixo}},
  \bibinfo{author}{\bibfnamefont{D.~A.} \bibnamefont{Lidar}}, \bibnamefont{and}
  \bibinfo{author}{\bibfnamefont{P.}~\bibnamefont{Zanardi}},
  \emph{\bibinfo{title}{Quantum adiabatic markovian master equations}},
  \bibinfo{journal}{New J. Phys.} \textbf{\bibinfo{volume}{14}},
  \bibinfo{pages}{123016} (\bibinfo{year}{2012}).

\bibitem[{\citenamefont{Smelyanskiy et~al.}(2017)\citenamefont{Smelyanskiy,
  Venturelli, Perdomo-Ortiz, Knysh, and Dykman}}]{Smelyanskiy_PRL2017}
\bibinfo{author}{\bibfnamefont{V.~N.} \bibnamefont{Smelyanskiy}},
  \bibinfo{author}{\bibfnamefont{D.}~\bibnamefont{Venturelli}},
  \bibinfo{author}{\bibfnamefont{A.}~\bibnamefont{Perdomo-Ortiz}},
  \bibinfo{author}{\bibfnamefont{S.}~\bibnamefont{Knysh}}, \bibnamefont{and}
  \bibinfo{author}{\bibfnamefont{M.~I.} \bibnamefont{Dykman}},
  \emph{\bibinfo{title}{Quantum annealing via environment-mediated quantum
  diffusion}}, \bibinfo{journal}{Phys. Rev. Lett.}
  \textbf{\bibinfo{volume}{118}}, \bibinfo{pages}{066802}
  (\bibinfo{year}{2017}).

\bibitem[{\citenamefont{Isakov and Moessner}(2003)}]{interplay}
\bibinfo{author}{\bibfnamefont{S.~V.} \bibnamefont{Isakov}} \bibnamefont{and}
  \bibinfo{author}{\bibfnamefont{R.}~\bibnamefont{Moessner}},
  \emph{\bibinfo{title}{Interplay of quantum and thermal fluctuations in a
  frustrated magnet}}, \bibinfo{journal}{Phys. Rev. B}
  \textbf{\bibinfo{volume}{68}}, \bibinfo{pages}{104409}
  (\bibinfo{year}{2003}).

\bibitem[{\citenamefont{Katzgraber}(2009)}]{katzgraber:09e}
\bibinfo{author}{\bibfnamefont{H.~G.} \bibnamefont{Katzgraber}},
  \emph{\bibinfo{title}{{{Introduction to Monte Carlo Methods}}}}
  (\bibinfo{year}{2009}), \bibinfo{note}{(arXiv:0905.1629)}.

\bibitem[{\citenamefont{Benedetti et~al.}(2017)\citenamefont{Benedetti,
  Realpe-G\'omez, Biswas, and Perdomo-Ortiz}}]{Benedetti2017}
\bibinfo{author}{\bibfnamefont{M.}~\bibnamefont{Benedetti}},
  \bibinfo{author}{\bibfnamefont{J.}~\bibnamefont{Realpe-G\'omez}},
  \bibinfo{author}{\bibfnamefont{R.}~\bibnamefont{Biswas}}, \bibnamefont{and}
  \bibinfo{author}{\bibfnamefont{A.}~\bibnamefont{Perdomo-Ortiz}},
  \emph{\bibinfo{title}{Quantum-assisted learning of hardware-embedded
  probabilistic graphical models}}, \bibinfo{journal}{Phys. Rev. X}
  \textbf{\bibinfo{volume}{7}}, \bibinfo{pages}{041052} (\bibinfo{year}{2017}).

\bibitem[{\citenamefont{Perdomo-Ortiz et~al.}(2018)\citenamefont{Perdomo-Ortiz,
  Benedetti, Realpe-G{\'o}mez, and Biswas}}]{PerdomoOrtiz2017}
\bibinfo{author}{\bibfnamefont{A.}~\bibnamefont{Perdomo-Ortiz}},
  \bibinfo{author}{\bibfnamefont{M.}~\bibnamefont{Benedetti}},
  \bibinfo{author}{\bibfnamefont{J.}~\bibnamefont{Realpe-G{\'o}mez}},
  \bibnamefont{and} \bibinfo{author}{\bibfnamefont{R.}~\bibnamefont{Biswas}},
  \emph{\bibinfo{title}{Opportunities and challenges for quantum-assisted
  machine learning in near-term quantum computers}}, \bibinfo{journal}{Quantum
  Science and Technology} \textbf{\bibinfo{volume}{3}}, \bibinfo{pages}{030502}
  (\bibinfo{year}{2018}).

\bibitem[{\citenamefont{Choi}(2008)}]{Choi2008}
\bibinfo{author}{\bibfnamefont{V.}~\bibnamefont{Choi}},
  \emph{\bibinfo{title}{Minor-embedding in adiabatic quantum computation: I.
  the parameter setting problem}}, \bibinfo{journal}{arXiv:0804.4884}
  (\bibinfo{year}{2008}).

\bibitem[{\citenamefont{Pudenz}(2016)}]{Pudenz2016}
\bibinfo{author}{\bibfnamefont{K.~L.} \bibnamefont{Pudenz}},
  \emph{\bibinfo{title}{Parameter setting for quantum annealers}},
  \bibinfo{journal}{arXiv:1611.07552}  (\bibinfo{year}{2016}).

\bibitem[{\citenamefont{Babbush et~al.}(2014)\citenamefont{Babbush,
  Perdomo-Ortiz, O'Gorman, Macready, and Aspuru-Guzik}}]{Babbush2014}
\bibinfo{author}{\bibfnamefont{R.}~\bibnamefont{Babbush}},
  \bibinfo{author}{\bibfnamefont{A.}~\bibnamefont{Perdomo-Ortiz}},
  \bibinfo{author}{\bibfnamefont{B.}~\bibnamefont{O'Gorman}},
  \bibinfo{author}{\bibfnamefont{W.}~\bibnamefont{Macready}}, \bibnamefont{and}
  \bibinfo{author}{\bibfnamefont{A.}~\bibnamefont{Aspuru-Guzik}},
  \emph{\bibinfo{title}{Construction of energy functions for lattice
  heteropolymer models: A case study in constraint satisfaction programming and
  adiabatic quantum optimization}}, \bibinfo{journal}{Adv. Chem. Phys.}
  \textbf{\bibinfo{volume}{155}}, \bibinfo{pages}{201} (\bibinfo{year}{2014}).

\bibitem[{\citenamefont{Babbush et~al.}(2013)\citenamefont{Babbush, O'Gorman,
  and Aspuru-Guzik}}]{babbush2013resource}
\bibinfo{author}{\bibfnamefont{R.}~\bibnamefont{Babbush}},
  \bibinfo{author}{\bibfnamefont{B.}~\bibnamefont{O'Gorman}}, \bibnamefont{and}
  \bibinfo{author}{\bibfnamefont{A.}~\bibnamefont{Aspuru-Guzik}},
  \emph{\bibinfo{title}{Resource efficient gadgets for compiling adiabatic
  quantum optimization problems}}, \bibinfo{journal}{Annalen der Physik}
  \textbf{\bibinfo{volume}{525}}, \bibinfo{pages}{877} (\bibinfo{year}{2013}).

\end{thebibliography}

\end{document}